\documentclass[10pt]{article}
\usepackage{graphicx}
\usepackage{amsmath}
\usepackage{amssymb}
\usepackage{caption2}
\begin{document}
\bibliographystyle{prsty}
\begin{center}
{\large {\bf \sc{  Analysis  of hadronic coupling constants  $ G_{B_c^*B_c\Upsilon}$, $ G_{B_c^*B_c J/\psi}$, $ G_{B_cB_c\Upsilon}$ and $ G_{B_cB_c J/\psi}$ with QCD sum rules }}} \\[2mm]
Zhi-Gang Wang \footnote{E-mail,zgwang@aliyun.com.  }     \\
 Department of Physics, North China Electric Power University,
Baoding 071003, P. R. China
\end{center}

\begin{abstract}
In this article, we study momentum dependence of the hadronic coupling constants $G_{B_c^* B_c \Upsilon}$, $G_{B_c^* B_c J/\psi}$, $G_{B_c B_c \Upsilon}$
and $G_{B_c B_c J/\psi}$ with  the off-shell $\Upsilon$ and $J/\psi$  using the three-point  QCD sum rules. Then we fit the  hadronic coupling constants  into
  analytical functions and extrapolate  them into  deep time-like regions to obtain the on-shell  values
  $ G_{B_c^*B_c\Upsilon}(q^2=M_{\Upsilon}^2)$, $ G_{B_c^*B_c J/\psi}(q^2=M_{J/\psi}^2)$, $ G_{B_cB_c\Upsilon}(q^2=M_{\Upsilon}^2)$ and $ G_{B_cB_c J/\psi}(q^2=M_{J/\psi}^2)$ for the first time. Those   hadronic coupling constants can be taken as basic input parameters in phenomenological analysis.
 \end{abstract}



\section{Introduction}
The suppression of  $J/\psi$ production in  relativistic  heavy ion
collisions is considered as   an important signature to identify the quark-gluon plasma
\cite{Matsui86}. The dissociation of $J/\psi$ in the quark-gluon
plasma due to color screening can lead to a reduction of its
production. The bottomonium states  are also sensitive to the color screening, the
 $\Upsilon$ suppression in high energy heavy ion collisions can also be taken  as a
signature to identify the quark-gluon plasma \cite{QGP-rev}.
The suppressions  on the
 $\Upsilon$ production in ultra-relativistic heavy ion collisions will be studied in details  at the Relativistic Heavy Ion Collider (RHIC) and
 Large Hadron Collider (LHC).
Before drawing a  definite conclusion on appearance  of the quark-gluon plasma, we have to  disentangle the color screening  versus recombination of the off-diagonal $\bar{c}c$ (or $\bar{b}b$ ) pairs in the hot dense medium  versus  cold nuclear matter  effects such as nuclear absorption,
shadowing and anti-shadowing \cite{Recombine-cc,CNM}.

We can  study the heavy quarkonium absorptions with the effective Lagrangians
in meson-exchange models \cite{meson-lagran}, and calculate the   absorption cross sections based on
  the interactions among the heavy quarkonia and heavy mesons, where the hadronic coupling constants are basic input parameters.
   The detailed knowledge  of the hadronic  coupling constants  is of great importance  in understanding  the effects of heavy quarkonium  absorptions in hadronic matter. Furthermore, the hadronic coupling constants among   the heavy quarkonia and heavy mesons
   play an important role in understanding
final-state interactions in the heavy quarkonium decays \cite{HeavyQ}.

The hadronic coupling constants in the
$D^*D \pi$,
$D^* D_s K$, $D_s^* D K$,
$B^*B \pi$,
$ B^{*}_{s}BK $,
$DD\rho$,
$D_{s}DK^*$, $ B_{s}BK^*$,
$D^{*}D \rho $, $D^{*}_{s}D K^{*}$, $B^{*}_{s}B K^{*}$,
$D^* D^* \rho$, $B^{*}B^{*}\rho $,
$B_{s0} B K$,  $B_{s1}B^{*}K$,
$ D^{*}_{s}D K_1 $, $ B^{*}_{s}BK_1 $,
$J/\psi D D$,
$J/\psi DD^*$,
$J/\psi D^*D^*$  vertices have been studied with the three-point QCD sum rules (QCDSR) \cite{3PTQCDSR,3PTQCDSR-Rev},
while the hadronic coupling constants in the
  $D^* D \pi$, $D^*D_sK$, $D^*_sDK$, $B^* B \pi$,
    $DD\rho$, $DD_sK^*$, $D_sD_s\phi$, $BB\rho$,
    $D^{*}D\rho$, $D^*D_s K^*$, $D_s^*D_s \phi$, $B^*B\rho$,
   $D^*D^*\pi$, $D^*D_s^*K$,  $B^*B^*\pi$,
   $D^* D^* \rho$,
   $D_0D\pi$, $B_0B\pi$,  $D_0D_sK$, $D_{s0}DK$,   $B_{s0}BK$,
      $D_1D^* \pi$,   $B_1B^*\pi$, $D_{s1} D^* K$, $B_{s1} B^* K$, $B_1 B_0 \pi$,
    $B_2B_1\pi$, $B_2B^*\pi$,
    $B_1B^*\rho$, $B_1B\rho$, $B_2B^*\rho$, $B_2B_1\rho$ vertices have been studied with
the  light-cone QCDSR \cite{LCQCDSR}.

 To my knowledge, the hadronic coupling constants among the heavy quarkonium states have not been studied with the three-point QCDSR or light-cone QCDSR. In the article, we study the vertices $ B_c^* B_c\Upsilon$, $ B_c^* B_c J/\psi$, $ B_c B_c\Upsilon$ and $ B_c B_c J/\psi$ with the three-point QCDSR.
 The QCD sum rules is a powerful nonperturbative approach   in
 studying   the heavy quarkonium states, and has given many successful descriptions of the masses, decay constants, form-factors, hadronic coupling constants  \cite{SVZ79,Reinders85,QCDSR-review}.

The $B_c^{*\pm}$ mesons have not been observed yet, but they are expected to be observed  at the LHC through the radiative transitions.
In previous works, we study the vector and axial-vector $B_c$ mesons with the QCDSR,
make reasonable predictions of  the masses and decay constants, then calculate the $B_c^* \to B_c$ electromagnetic form-factor with the three-point QCDSR, and obtain the decay width of the radiative transitions $B_c^{*\pm} \to B_c^{\pm}\gamma$ \cite{Wang-BC,Wang2013}.

The article is arranged as follows:  we study the $ B_c^* B_c\Upsilon$, $ B_c^* B_c J/\psi$, $ B_c B_c\Upsilon$ and $ B_c B_c J/\psi$ vertices using
  the three-point QCDSR in Sect.2; in Sect.3, we present the numerical results and discussions; and Sect.4 is reserved for our
conclusions.

\section{ The $B_c^* B_c \Upsilon $ and  $B_c B_c \Upsilon $ (also  $B_c^* B_c J/\psi $ and $B_c B_c J/\psi $) vertices with QCD sum rules}
We study the $B_c^* B_c \Upsilon $ and $B_c B_c \Upsilon $ vertices with  the three-point correlation functions $\Pi_{\mu\nu}(p,p^{\prime})$ and $\Pi_{\mu}(p,p^{\prime})$,  respectively,
\begin{eqnarray}
\Pi_{\mu\nu}(p,p^{\prime})&=&i^2\int d^4 x d^4 y e^{ip^{\prime}\cdot x+i(p-p^{\prime})\cdot y} \langle 0|T\left\{J_5(x) j_\mu (y) J_\nu^{\dagger}(0)\right\} |0\rangle \, , \nonumber\\
\Pi_{\mu}(p,p^{\prime})&=&i^2\int d^4 x d^4 y e^{ip^{\prime}\cdot x+i(p-p^{\prime})\cdot y} \langle 0|T\left\{J_5(x) j_\mu (y) J_5^{\dagger}(0)\right\} |0\rangle \, ,
\end{eqnarray}
where
\begin{eqnarray}
j_\mu(x)&=& \bar{b}(x)\gamma_\mu b(x) \, , \nonumber\\
J_5(x)&=&\bar{c}(x)i\gamma_5b(x)\, , \nonumber\\
J^{\dagger}_\nu(x)&=&\bar{b}(x)\gamma_\nu c(x)\, ,
\end{eqnarray}
the currents $j_\mu(x)$, $J_5(x)$ and  $J^{\dagger}_\nu(x)$ interpolate the heavy quarkonia $\Upsilon$,  $B_c$ and $B_c^*$, respectively.

We can insert  a complete set of intermediate hadronic states with
the same quantum numbers as the current operators $j_{\mu}(y)$, $J_5(x)$, $J^{\dagger}_\nu(0)$ and $J^{\dagger}_5(0)$ into the
correlation functions $\Pi_{\mu\nu}(p,p^{\prime})$ and $\Pi_{\mu}(p,p^{\prime})$  to obtain the hadronic representation
\cite{SVZ79,Reinders85}. After isolating the ground state
contributions come from the  heavy quarkonia $\Upsilon$,  $B_c$ and $B_c^*$, we get the following results,
\begin{eqnarray}
\Pi_{\mu\nu}(p,p^{\prime})&=&\frac{\langle 0|J_5(0) |B_c(p^\prime)\rangle \langle 0|j_\mu(0)|\Upsilon(q)\rangle  \langle B_c^*(p)  |J^{\dagger}_\nu(0) |0   \rangle
\langle B_c(p^\prime) \Upsilon(q) |{ \mathcal{L}}(0)|B_c^*(p)\rangle}{(M_{B_c}^2-p^{\prime2})(M_{\Upsilon}^2-q^2)(M_{B_c^*}^2-p^2)} +\cdots   \, ,\nonumber\\
&=&-\frac{f_{B_c}M_{B_c}^2f_{B_c^*}M_{B_c^*}f_{\Upsilon} M_{\Upsilon} }{(m_b+m_c)(M_{B_c}^2-p^{\prime2})(M_{B_c^*}^2-p^2)(M_{\Upsilon}^2-q^2)}  \,G_{B_c^* B_c \Upsilon}(q^2)\,\epsilon_{\mu\nu\alpha\beta}p^\alpha p^{\prime\beta} +\cdots \, , \\
\Pi_{\mu}(p,p^{\prime})&=&\frac{\langle 0|J_5(0) |B_c(p^\prime)\rangle \langle 0|j_\mu(0)|\Upsilon(q)\rangle  \langle B_c(p)  |J^{\dagger}_5(0) |0   \rangle
\langle B_c(p^\prime) \Upsilon(q) |{ \mathcal{L}}(0)|B_c(p)\rangle}{(M_{B_c}^2-p^{\prime2})(M_{\Upsilon}^2-q^2)(M_{B_c}^2-p^2)} +\cdots   \, ,\nonumber\\
&=&\frac{f_{B_c}^2M_{B_c}^4 f_{\Upsilon} M_{\Upsilon} }{(m_b+m_c)^2(M_{B_c}^2-p^{\prime2})(M_{B_c}^2-p^2)(M_{\Upsilon}^2-q^2)}  \,G_{B_c B_c \Upsilon}(q^2)\,(p+p^{\prime})_{\mu }  +\cdots \, , \nonumber\\
&=&\Gamma_p(p,p^{\prime}) p_\mu +\Gamma_{p^{\prime}}(p,p^{\prime}) p^{\prime}_\mu+\cdots \, ,
\end{eqnarray}
where we have used the following  effective Lagrangian ${\mathcal{L}}$ and definitions for  the decay constants $f_{B_c^*}$, $f_{B_c}$, $f_{\Upsilon}$,
\begin{eqnarray}
{ \mathcal{L}}&=& G_{B_c^* B_c \Upsilon}\, \epsilon_{\lambda\tau\rho\sigma}B_c^{\dagger}\partial^{\lambda}\Upsilon^\tau \partial^{\rho}B_c^{*\sigma}+iG_{B_c B_c J/\psi}\Upsilon^\mu \left(B_c^{\dagger}\partial_\mu B_c-\partial_\mu B_c^{\dagger} B_c\right)\, ,
\end{eqnarray}
\begin{eqnarray}
\langle0|J_\mu(0) |B_c^*(p)\rangle&=&f_{B_c^*}M_{B_c^*}\zeta_\mu \, ,\nonumber\\
\langle0|J_5(0) |B_c(p^{\prime})\rangle&=&\frac{f_{B_c}M_{B_c}^2}{m_b+m_c} \, ,\nonumber\\
\langle0|j_\mu(0) |\Upsilon(q)\rangle&=&f_{\Upsilon}M_{\Upsilon}\xi_\mu \, ,
\end{eqnarray}
$q_\mu=(p-p^{\prime})_\mu$, the $\zeta_\mu$ and $\xi_\mu$ are the polarization  vectors.
The tensor structures $p_\mu$ and $p^{\prime}_{\mu}$ associate with the correlation functions $\Gamma_{p}(p,p^{\prime})$  and $\Gamma_{p^{\prime}}(p,p^{\prime})$,  respectively, we obtain the QCDSR  by considering the combination $\Gamma_{p}(p,p^{\prime})+\Gamma_{p^{\prime}}(p,p^{\prime})$.

The effective fields describe
point-like particles only in the case that all the interacting particles are on the mass-shell. When at least one
  particle  in the vertex is off-shell, the finite-size effects of the hadrons become important.
We should introduce form-factors in the hadronic coupling constants to parameterize the off-shell  effects, which are of great importance in
calculating scattering amplitudes at the hadronic level. In this article, we parameterize  the $q^2$ dependence of the hadronic coupling constants $G(q^2)$ with suitable functions, then obtain
the on-shell  values $G(q^2=-Q^2=M_{\Upsilon}^2)$ by analytically continuing the $q^2$ to the physical region.

Now, we briefly outline  the operator product expansion for the correlation functions $\Pi_{\mu\nu}(p,p^{\prime})$ and $\Pi_{\mu}(p,p^{\prime})$.  We contract the quark fields in the correlation functions $\Pi_{\mu\nu}(p,p^{\prime})$ and $\Pi_{\mu}(p,p^{\prime})$ with Wick theorem firstly,
\begin{eqnarray}
\Pi_{\mu\nu}(p,p^{\prime})&=&\int d^4 x d^4 y e^{ip^{\prime}\cdot x+i(p-p^{\prime})\cdot y} \,\,{\rm Tr} \left\{i\gamma_5B^{mn}(x-y)\gamma_{\mu}B^{nk}(y)\gamma_\nu C^{km}(-x) \right\}  \, , \nonumber\\
\Pi_{\mu}(p,p^{\prime})&=&\int d^4 x d^4 y e^{ip^{\prime}\cdot x+i(p-p^{\prime})\cdot y} \,\,{\rm Tr} \left\{i\gamma_5B^{mn}(x-y)\gamma_{\mu}B^{nk}(y)i\gamma_5 C^{km}(-x) \right\}  \, ,
\end{eqnarray}
replace the $b$ and $c$ quark propagators $B^{ij}(x) $ and $C^{ij}(x)$ with the corresponding full propagators $S_{ij}(x)$,
\begin{eqnarray}
S_{ij}(x)&=&\frac{i}{(2\pi)^4}\int d^4k e^{-ik \cdot x} \left\{
\frac{\delta_{ij}}{\!\not\!{k}-m_Q}
-\frac{g_sG^n_{\alpha\beta}t^n_{ij}}{4}\frac{\sigma^{\alpha\beta}(\!\not\!{k}+m_Q)+(\!\not\!{k}+m_Q)
\sigma^{\alpha\beta}}{(k^2-m_Q^2)^2}+\frac{\delta_{ij}\langle g^2_sGG\rangle }{12}\right.\nonumber\\
&&\left. \frac{m_Qk^2+m_Q^2\!\not\!{k}}{(k^2-m_Q^2)^4}+\cdots\right\} \, ,
\end{eqnarray}
 where $Q=c,b$, $\langle g^2_sGG\rangle=\langle g^2_sG^n_{\alpha\beta}G^{n\alpha\beta}\rangle$, $t^n=\frac{\lambda^n}{2}$, the $\lambda^n$ are the Gell-Mann matrixes, the $i$, $j$ are color indexes \cite{Reinders85},
then compute  the  integrals. In this article, we take into account the leading-order   contributions $\Pi_{\mu\nu}^{0}(p,p^{\prime})$, $\Pi_{\mu}^{0}(p,p^{\prime})$ and  gluon condensate contributions $\Pi_{\mu\nu}^{GG}(p,p^{\prime})$, $\Pi_{\mu}^{GG}(p,p^{\prime})$ in the operator product expansion, and show them explicitly using the Feynman diagrams in Figs.1-2.

The leading-order   contributions $\Pi_{\mu\nu}^{0}(p,p^{\prime})$, $\Pi_{\mu}^{0}(p,p^{\prime})$ can be written as
\begin{eqnarray}
\Pi_{\mu\nu}^{0}(p,p^{\prime})&=&\frac{3}{(2\pi)^4}\int d^4k \frac{ {\rm Tr}\left\{ \gamma_5\left[ \!\not\!{k}+m_b\right]\gamma_\mu \left[ \!\not\!{k}+\!\not\!{p} -\!\not\!{p^{\prime}}+ m_b\right]\gamma_\nu\left[ \!\not\!{k}-\!\not\!{p^{\prime}} +m_c\right]\right\}}{\left[k^2-m_b^2\right]\left[(k+p-p^{\prime})^2-m_b^2\right]\left[(k-p^{\prime})^2-m_c^2\right]}\, ,\nonumber\\
&=&\int ds du \frac{\rho_{\mu\nu}(s,u,q^2)}{(s-p^2)(u-p^{\prime2})} \, ,\\
\Pi_{\mu}^{0}(p,p^{\prime})&=&\frac{3i}{(2\pi)^4}\int d^4k \frac{ {\rm Tr}\left\{ \gamma_5\left[ \!\not\!{k}+m_b\right]\gamma_\mu \left[ \!\not\!{k}+\!\not\!{p} -\!\not\!{p^{\prime}}+ m_b\right]\gamma_5\left[ \!\not\!{k}-\!\not\!{p^{\prime}} +m_c\right]\right\}}{\left[k^2-m_b^2\right]\left[(k+p-p^{\prime})^2-m_b^2\right]\left[(k-p^{\prime})^2-m_c^2\right]}\, ,\nonumber\\
&=&\int ds du \frac{\rho_{\mu}(s,u,q^2)}{(s-p^2)(u-p^{\prime2})} \, .
\end{eqnarray}
We put all the quark lines on mass-shell using the Cutkosky's rules, see Fig.1,
and  obtain the leading-order  spectral densities  $\rho_{\mu\nu}(s,u,q^2)$ and $\rho_{\mu}(s,u,q^2)$,
\begin{eqnarray}
\rho_{\mu\nu}(s,u,q^2)  &=&-\frac{3i}{(2\pi)^3} \int d^4k \delta\left[k^2-m_b^2\right]\delta\left[(k+p-p^{\prime})^2-m_b^2\right]\delta\left[(k-p^{\prime})^2-m_c^2\right]\nonumber\\
&& {\rm Tr}\left\{ \gamma_5\left[ \!\not\!{k}+m_b\right]\gamma_\mu \left[ \!\not\!{k}+\!\not\!{p} -\!\not\!{p^{\prime}}+ m_b\right]\gamma_\nu\left[ \!\not\!{k}-\!\not\!{p^{\prime}} +m_c\right]\right\} \nonumber\\
&=& -\frac{3 \epsilon_{\mu\nu\alpha\beta}p^\alpha p^{\prime\beta}}{ 4\pi^2\sqrt{\lambda(s,u,q^2)}}\left\{m_b +\frac{(m_b-m_c)(s+u-q^2+2m_b^2-2m_c^2)q^2}{\lambda(s,u,q^2)}\right\}   \, ,
\end{eqnarray}
\begin{eqnarray}
\rho_{\mu}(s,u,q^2)  &=&\frac{3}{(2\pi)^3} \int d^4k \delta\left[k^2-m_b^2\right]\delta\left[(k+p-p^{\prime})^2-m_b^2\right]\delta\left[(k-p^{\prime})^2-m_c^2\right]\nonumber\\
&& {\rm Tr}\left\{ \gamma_5\left[ \!\not\!{k}+m_b\right]\gamma_\mu \left[ \!\not\!{k}+\!\not\!{p} -\!\not\!{p^{\prime}}+ m_b\right]\gamma_5\left[ \!\not\!{k}-\!\not\!{p^{\prime}} +m_c\right]\right\} \nonumber\\
&=& \frac{3 }{ 8\pi^2\sqrt{\lambda(s,u,q^2)}}  \left\{  \left[ s+u-q^2-2(m_b-m_c)^2\right](C_p p_\mu+C_{p^\prime}p^{\prime}_\mu)\right.\nonumber\\
&&\left.+\left[u-(m_b-m_c)^2 \right]q_\mu+q^2 p^{\prime}_\mu\right\} \, ,
\end{eqnarray}
where
\begin{eqnarray}
C_{p}&=&\frac{(s+u-q^2)(u+m_b^2-m_c^2)-2u(u-q^2+m_b^2-m_c^2)}{\lambda(s,u,q^2)} \, , \nonumber\\
C_{p^\prime}&=&\frac{(s+u-q^2)(u-q^2+m_b^2-m_c^2 )-2s(u+m_b^2-m_c^2)}{\lambda(s,u,q^2)} \, ,
\end{eqnarray}
and $\lambda(a,b,c)=a^2+b^2+c^2-2ab-2bc-2ca$,  we have  used the formulae presented in Refs.\cite{Wang1209,Gongshi-Ioffe}
to compute  the integrals.

\begin{figure}
 \centering
 \includegraphics[totalheight=6cm,width=7cm]{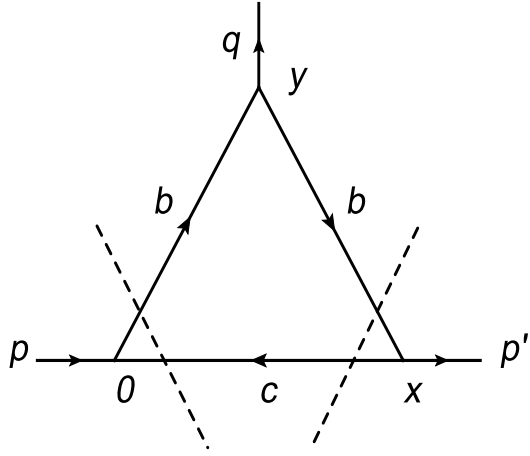}
    \caption{The leading-order   contributions, the dashed lines denote the Cutkosky's cuts. }
\end{figure}

\begin{figure}
 \centering
 \includegraphics[totalheight=8cm,width=14cm]{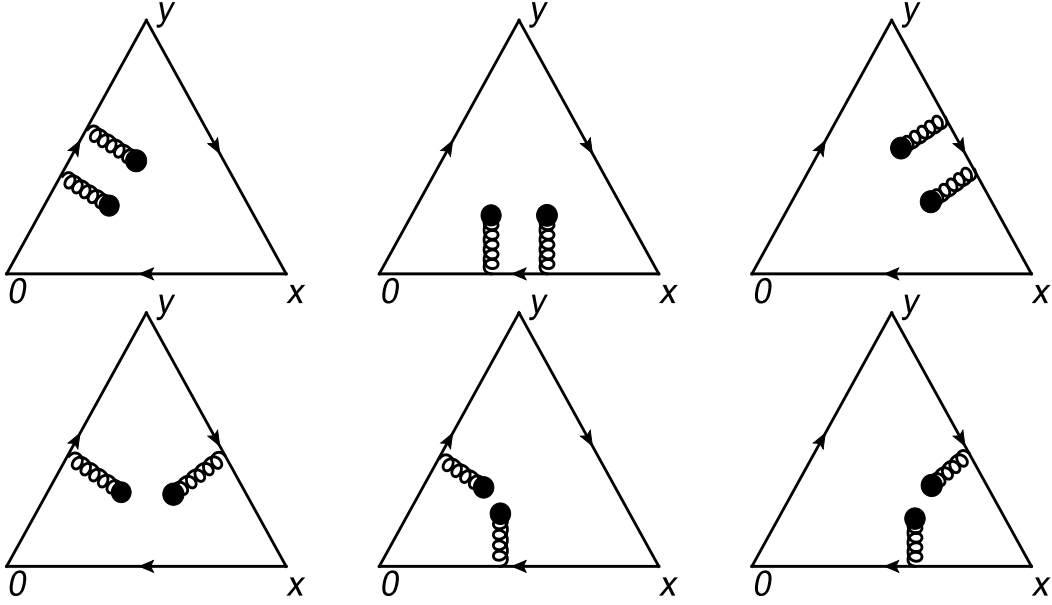}
    \caption{The gluon condensate   contributions. }
\end{figure}

 We calculate the gluon condensate contributions directly (see Fig.2) and obtain the following formulas,
\begin{eqnarray}
\Pi_{\mu\nu}^{GG}(p,p^{\prime})&=&\frac{i\epsilon_{\mu\nu\alpha\beta}}{4\pi^2} \langle\frac{\alpha_sGG}{\pi}\rangle \left\{-m_b^3 \left(\overline{I}_{411}+\overline{I}_{141}\right)p^{\alpha} p^{\prime\beta} -m_b^2(m_b-m_c)\left(\overline{I}^\alpha_{411}+\overline{I}^\alpha_{141}\right)q^\beta\right.\nonumber\\
&&-m_b m_c^2\overline{I}_{114}p^\alpha p^{\prime\beta}+m_c^2(m_c-m_b)\overline{I}^\alpha_{114}q^\beta
-m_b\overline{I}_{311}p^\alpha p^{\prime\beta}-m_b\overline{I}^\alpha_{311}p^\beta\nonumber\\
&&\left.+m_b\overline{I}^\alpha_{131}p^{\prime\beta}+m_c\overline{I}_{113}^\alpha q^\beta \right\}\nonumber\\
&&+\frac{i\epsilon_{\mu\nu\alpha\beta}}{24\pi^2} \langle\frac{\alpha_sGG}{\pi}\rangle \left\{ (m_b-m_c)\left(\overline{I}_{221}^\alpha+\overline{I}_{122}^\alpha\right) q^\beta +m_b\left(\overline{I}_{221}+\overline{I}_{122}\right)p^\alpha p^{\prime\beta}\right. \nonumber\\
&&\left. -2m_b\overline{I}_{212}^\alpha p^{\prime\beta}-3(m_b-m_c)\overline{I}_{212}^\alpha q^\beta-3m_b \overline{I}_{212} p^\alpha p^{\prime\beta} \right\} \, ,
\end{eqnarray}

\begin{eqnarray}
\Pi^{\mu}_{GG}(p,p^{\prime})&=&\frac{i}{8\pi^2} \langle\frac{\alpha_sGG}{\pi}\rangle \left\{ m_b^2\left[ -\left(2(m_b-m_c)^2+q^2\right)\left(\overline{I}_{411}^\mu+\overline{I}_{141}^\mu\right)+\overline{K}^\mu_{411}+\overline{K}^\mu_{141}+\overline{N}^\mu_{411}\right.\right.\nonumber\\
&&\left.+\overline{N}^\mu_{141}-(m_b-m_c)^2\left(\overline{I}_{411}+\overline{I}_{141}\right)q^\mu+\left(\overline{N}_{411}+\overline{N}_{141}\right)q^\mu
+q^2\left(\overline{I}_{411}+\overline{I}_{141}\right) p^{\prime\mu}\right] \nonumber\\
&&\left.+m_c^2\left[ -\left(2(m_b-m_c)^2+q^2\right)\overline{I}_{114}^\mu+\overline{K}^\mu_{114}+\overline{N}^\mu_{114}-(m_b-m_c)^2\overline{I}_{114}q^\mu+\overline{N}_{114}q^\mu\right.\right.\nonumber\\
&&\left.+q^2\overline{I}_{114} p^{\prime\mu}\right] +m_b\left[2(m_c-m_b)\left(\overline{I}^\mu_{311}+\overline{I}^\mu_{131}\right)+m_b\left(\overline{I}_{311}+\overline{I}_{131}\right)p^{\prime\mu} \right.\nonumber\\
&&\left.\left. +(2m_c-m_b)\overline{I}_{311}q^\mu\right]+m_c\left[ (2m_b-m_c)\left(2\overline{I}^\mu_{113}+\overline{I}_{113}q^\mu \right)\right] \right\} \nonumber\\
&&+\frac{i}{48\pi^2} \langle\frac{\alpha_sGG}{\pi}\rangle \left\{ \left(12m_bm_c-4m_b^2-2m_c^2+q^2\right)\overline{I}^\mu_{221}+\overline{K}_{221}^\mu+\overline{N}_{221}^\mu \right. \nonumber \\
&&+\left[\left(6m_bm_c-m_b^2-m_c^2\right)\left(\overline{I}_{221}+\overline{I}_{122}\right)+\overline{N}_{221}+\overline{N}_{122}\right]q^\mu+\left(2m_b^2-q^2\right)\overline{I}_{221}p^{\prime\mu} \nonumber\\
&&+\left(18m_bm_c-8m_b^2-4m_c^2-q^2\right)\left(\overline{I}_{122}^\mu+\overline{I}_{212}^\mu\right)+3\overline{K}_{122}^\mu+\overline{N}_{122}^\mu+\overline{K}_{212}^\mu+3\overline{N}_{212}^\mu\nonumber\\
&&+\left(4m_b^2+q^2\right)\left(\overline{I}_{122}+\overline{I}_{212}\right)p^{\prime\mu}+3\left(4m_bm_c-m_b^2-m_c^2\right)\overline{I}_{212}q^\mu+3\overline{N}_{212}q^\mu \nonumber\\
&&-\left(2\overline{I}_{121}+3\overline{I}_{112}+3\overline{I}_{211}\right)q^\mu+\left(\overline{I}_{121}-2\overline{I}_{112}+\overline{I}_{211}\right)p^{\prime\mu}
\nonumber\\
&&\left.-\left(6\overline{I}^\mu_{121}+4\overline{I}^\mu_{112}+6\overline{I}^\mu_{211}\right) \right\} \, ,
\end{eqnarray}

where
\begin{eqnarray}
\overline{I}_{ijn}&=& \int d^4k \frac{1}{\left[k^2-m_b^2\right]^i \left[(k+p-p^{\prime})^2-m_b^2\right]^j \left[ (k-p^{\prime})^2-m_c^2\right]^n} \, , \nonumber\\
\overline{K}_{ijn}&=& \int d^4k \frac{p^2}{\left[k^2-m_b^2\right]^i \left[(k+p-p^{\prime})^2-m_b^2\right]^j \left[ (k-p^{\prime})^2-m_c^2\right]^n} \, , \nonumber\\
\overline{N}_{ijn}&=& \int d^4k \frac{{p^\prime}^2}{\left[k^2-m_b^2\right]^i \left[(k+p-p^{\prime})^2-m_b^2\right]^j \left[ (k-p^{\prime})^2-m_c^2\right]^n} \, , \nonumber\\
\overline{I}_{ijn}^\alpha&=& \int d^4k \frac{k^\alpha}{\left[k^2-m_b^2\right]^i \left[(k+p-p^{\prime})^2-m_b^2\right]^j \left[ (k-p^{\prime})^2-m_c^2\right]^n}  \, , \nonumber\\
\overline{K}_{ijn}^\alpha&=& \int d^4k \frac{p^2 k^\alpha}{\left[k^2-m_b^2\right]^i \left[(k+p-p^{\prime})^2-m_b^2\right]^j \left[ (k-p^{\prime})^2-m_c^2\right]^n}  \, , \nonumber\\
\overline{N}_{ijn}^\alpha&=& \int d^4k \frac{{p^\prime}^2 k^\alpha}{\left[k^2-m_b^2\right]^i \left[(k+p-p^{\prime})^2-m_b^2\right]^j \left[ (k-p^{\prime})^2-m_c^2\right]^n}  \, .
\end{eqnarray}

 We take  quark-hadron duality  below the thresholds
$s_0$ and $u_0$ for the mesons $B_c^*$ (or $B_c$) and $B_c$, respectively,
 perform double Borel transform  with respect to the variables
$P^2=-p^2$ and $P^{\prime2}=-p^{\prime2}$, respectively,
and  obtain the QCDSR   for the coupling constants $G_{B_c^* B_c \Upsilon}(q^2)$ and $G_{B_c B_c \Upsilon}(q^2)$,
 \begin{eqnarray}
 G_{B_c^* B_c \Upsilon}(q^2)&=&\frac{(m_b+m_c)(M_{\Upsilon}^2-q^2)}{f_{\Upsilon}f_{B_c^*}f_{B_c}M_{\Upsilon}M_{B_c^*}M_{B_c}^2}\exp\left( \frac{M_{B_c^*}^2}{M_1^2}+\frac{M_{B_c}^2}{M_2^2}\right)\nonumber\\
 && \frac{3}{ 4\pi^2}\int_{(m_b+m_c)^2}^{s_0} ds \int_{(m_b+m_c)^2}^{u_0} du \frac{\mathcal{C}}{\sqrt{\lambda(s,u,q^2)}}\exp\left( -\frac{s}{M_1^2}-\frac{u}{M_2^2}\right) \nonumber\\
&&\left\{m_b +\frac{(m_b-m_c)(s+u-q^2+2m_b^2-2m_c^2)q^2}{\lambda(s,u,q^2)}\right\}|_{|f(s,u,q^2)|\leq 1} \nonumber\\
&&-\frac{(m_b+m_c)(M_{\Upsilon}^2-q^2)M_1^2M_2^2}{f_{\Upsilon}f_{B_c^*}f_{B_c}M_{\Upsilon}M_{B_c^*}M_{B_c}^2}\langle\frac{\alpha_sGG}{\pi}\rangle  \exp\left( \frac{M_{B_c^*}^2}{M_1^2}+\frac{M_{B_c}^2}{M_2^2}\right) \nonumber\\
&&\left\{ \frac{ m_b^3}{4\pi^2} \left(I_0^{411}+I_0^{141}\right)-\frac{ m_b^2(m_b-m_c)}{4\pi^2}\left(I_{01}^{411}+I_{01}^{141} \right) + \frac{m_b m_c^2}{4\pi^2}  I_0^{114}    \right. \nonumber\\
&&   +\frac{m_c^2(m_c-m_b)}{4\pi^2} I_{01}^{114}+\frac{ m_b}{4\pi^2}\left(I_{0}^{311}-I_{01}^{311}+I_{10}^{311} \right)-\frac{m_b}{4\pi^2}I_{10}^{131}+\frac{m_c}{4\pi^2}I_{01}^{113} +\frac{m_b}{8\pi^2}I_{0}^{212} \nonumber\\
&&\left.+\frac{m_b-m_c}{24\pi^2}\left(  I_{01}^{221} +I_{01}^{122}\right)-\frac{m_b}{24\pi^2}\left(I_{0}^{221}+I_{0}^{122}\right)+\frac{m_b}{12\pi^2}I_{10}^{212}-\frac{ m_b-m_c}{8\pi^2}I_{01}^{212}\right\}\, , \nonumber\\
\end{eqnarray}

 \begin{eqnarray}
 G_{B_c B_c \Upsilon}(q^2)&=&\frac{(m_b+m_c)^2(M_{\Upsilon}^2-q^2)}{2f_{\Upsilon}f_{B_c}^2M_{\Upsilon}M_{B_c}^4}\exp\left(\frac{M_{B_c}^2}{M_1^2}+\frac{M_{B_c}^2}{M_2^2}\right)\nonumber\\
 && \frac{3}{ 8\pi^2}\int_{(m_b+m_c)^2}^{s_0} ds \int_{(m_b+m_c)^2}^{u_0} du \frac{\mathcal{C}}{\sqrt{\lambda(s,u,q^2)}}\exp\left( -\frac{s}{M_1^2}-\frac{u}{M_2^2}\right) \nonumber\\
&&\left\{\left[ s+u-q^2-2(m_b-m_c)^2\right]\left[\frac{(u-s-q^2)(u+m_b^2-m_c^2)}{\lambda(s,u,q^2)}   \right.\right.\nonumber\\
&&\left.\left.+\frac{(s-u-q^2)(u-q^2+m_b^2-m_c^2 )}{\lambda(s,u,q^2)}\right]+q^2\right\}|_{|f(s,u,q^2)|\leq 1} \nonumber\\
&&-\frac{(m_b+m_c)^2(M_{\Upsilon}^2-q^2)M_1^2M_2^2}{ 2f_{\Upsilon}f_{B_c}^2M_{\Upsilon}M_{B_c}^4}\langle\frac{\alpha_sGG}{\pi}\rangle  \exp\left(\frac{M_{B_c}^2}{M_1^2}+\frac{M_{B_c}^2}{M_2^2}\right) \nonumber\\
&&\left\{\frac{m_b^2}{8\pi^2}\left[ -\left(2(m_b-m_c)^2+q^2\right)\left(I_{01}^{411}+I_{01}^{141}\right)+K_{01}^{411}+K_{01}^{141}+N_{01}^{411}+N_{01}^{141}\right. \right. \nonumber\\
&&\left. +q^2\left(I_{0}^{411}+I_{0}^{141}\right)\right]+\frac{m_c^2}{8\pi^2}\left[-\left(2(m_b-m_c)^2+q^2 \right)I_{01}^{114}+K_{01}^{114}+N_{01}^{114}+q^2I_{0}^{114} \right]\nonumber\\
&& + \frac{ 2m_b(m_c-m_b)\left(I_{01}^{311}+I_{01}^{131}\right)+m_b^2\left(I_{0}^{311}+I_{0}^{131}  \right)
 +2m_c(2m_b-m_c)I_{01}^{113}}{8\pi^2}    \nonumber\\
 &&+\frac{12m_bm_c-4m_b^2-2m_c^2+q^2}{48\pi^2}I_{01}^{221}+\frac{18m_bm_c-8m_b^2-4m_c^2-q^2}{48\pi^2}\left(I_{01}^{122}+I_{01}^{212}\right) \nonumber\\
 &&+\frac{K_{01}^{221}+N_{01}^{221}+3K_{01}^{122}+N_{01}^{122}+K_{01}^{212}+3N_{01}^{212}+(2m_b^2-q^2)I_{0}^{221}}{48\pi^2} \nonumber\\
 && \left.+\frac{(4m_b^2+q^2)\left(I_{0}^{122}+I_{0}^{212} \right)}{48\pi^2}-\frac{3I_{01}^{121}+2I_{01}^{112}+3I_{01}^{211}}{24\pi^2}+\frac{I_{0}^{121}-2I_{0}^{112}+I_{0}^{211}}{48\pi^2}\right\} \, ,\nonumber\\
\end{eqnarray}
where
\begin{eqnarray}
f(s,u,q^2) &=& \frac{(s+u-q^2)(u-q^2+m_b^2-m_c^2)-2s(u+m_b^2-m_c^2)}{\sqrt{\lambda(s,u,q^2)\left[(u-q^2+m_b^2-m_c^2)^2-4sm_b^2\right]}}  \, , \nonumber\\
{\mathcal{C}}&=&\sqrt{\frac{4\pi\alpha_s^\mathcal{C}}{3v_s} \left[1-\exp\left(-\frac{4\pi\alpha_s^\mathcal{C}}{3v_s}\right)\right]^{-1}}\sqrt{\frac{4\pi\alpha_s^\mathcal{C}}{3v_u} \left[1-\exp\left(-\frac{4\pi\alpha_s^\mathcal{C}}{3v_u}\right)\right]^{-1}}\, ,\nonumber\\
v_s&=&\sqrt{1-\frac{4m_bm_c}{s-(m_b-m_c)^2}}\, ,\nonumber\\
v_u&=&\sqrt{1-\frac{4m_bm_c}{u-(m_b-m_c)^2}}\, ,
\end{eqnarray}
the explicit expressions of the $I_0^{ijn}$, $K_0^{ijn}$, $N_0^{ijn}$, $I_{10}^{ijn}$, $I_{01}^{ijn}$, $K_{10}^{ijn}$, $K_{01}^{ijn}$, $N_{10}^{ijn}$, $N_{01}^{ijn}$ are presented in the appendix.
For the heavy quarkonium states $B_c^*$ and $B_c$, the relative velocities of quark movement are small, we should account for the Coulomb-like $\frac{\alpha_s^\mathcal{C}}{v_s}$ and  $\frac{\alpha_s^\mathcal{C}}{v_u}$ corrections correspond to the  currents $J^{\dagger}_\nu(0)$ (or $J^{\dagger}_5(0)$) and $J_5(x)$, respectively.  After taking into account  all the Coulomb-like contributions shown in Fig.3, we obtain the coefficient $\mathcal{C}$ to dress the   quark-meson  vertexes \cite{Kiselev2000,Coulomb-BC}, and take the approximation ${\alpha_s^\mathcal{C}}=\alpha_s(\mu)$ in numerical  calculations \cite{Wang-BC}.

We can obtain the hadronic coupling constants $G_{B_c^*B_c J/\psi}(q^2)$ and $G_{B_cB_c J/\psi}(q^2)$ with the following simple replacements,
\begin{eqnarray}
G_{B_c^*B_c J/\psi}(q^2)&=&G_{B_c^*B_c \Upsilon}(q^2)|_{m_b\leftrightarrow m_c, \,f_{\Upsilon}\rightarrow f_{J/\psi}, \,M_{\Upsilon}\rightarrow M_{J/\psi}} \, ,\nonumber\\
G_{B_cB_c J/\psi}(q^2)&=&G_{B_cB_c \Upsilon}(q^2)|_{m_b\leftrightarrow m_c, \,f_{\Upsilon}\rightarrow f_{J/\psi}, \,M_{\Upsilon}\rightarrow M_{J/\psi}} \, .
\end{eqnarray}

\begin{figure}
 \centering
 \includegraphics[totalheight=3.5cm,width=6cm]{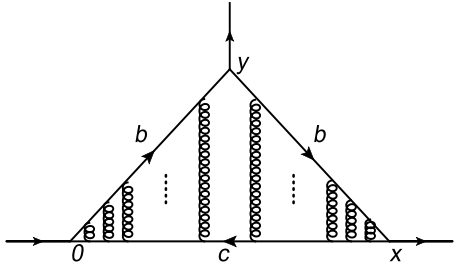}
      \caption{The ladder Feynman diagram for the Coulomb-like interactions. }
\end{figure}

In this article, we calculate the hadronic coupling constants $G_{B_c^*B_c \Upsilon}$, $G_{B_cB_c \Upsilon}$, $G_{B_c^*B_c J/\psi}$ $G_{B_cB_c J/\psi}$
at the space-like region $Q^2=-q^2\geq 1\,\rm{GeV}^2$, then fit the $G_{B_c^*B_c \Upsilon}$, $G_{B_cB_c \Upsilon}$, $G_{B_c^*B_c J/\psi}$ $G_{B_cB_c J/\psi}$  into  suitable analytical functions, and obtain the values $G_{B_c^*B_c \Upsilon}(q^2=M_{\Upsilon}^2)$, $G_{B_cB_c \Upsilon}(q^2=M_{\Upsilon}^2)$,  $G_{B_c^*B_c J/\psi}(q^2=M_{J/\psi}^2)$ and $G_{B_cB_c J/\psi}(q^2=M_{J/\psi}^2)$ by analytically  continuing  the variable $q^2$ to the physical regions.

\section{Numerical results and discussions}
The hadronic input parameters are taken as $f_{B_c^*}=0.384 \,\rm{GeV}$, $M_{B_c^*}=
6.337 \,\rm{GeV}$ from the QCDSR \cite{Wang-BC}, $f_{B_c}=395 \rm{MeV}$ from the QCD-motivated potential model \cite{Kiselev04}, $M_{B_c}=6.277\,\rm{GeV}$, $M_{\Upsilon}=9.4603\,\rm{GeV}$, $M_{J/\psi}=3.096916\,\rm{GeV}$ from the Particle Data Group \cite{PDG}. We extract the values of the decay constants $f_{\Upsilon}=0.700\,\rm{GeV}$ and $ f_{J/\psi}=0.415\,\rm{GeV}$ from the decays $\Upsilon\to e^+ e^-$ and $J/\psi\to e^+ e^-$, respectively \cite{PDG}.
The decay constants have the relation $f_{B_c^*} \approx f_{B_c}$,   the masses have the splitting $M_{B_c^*}-M_{B_c}=60\,\rm{MeV}$.  The calculations based on the nonrelativistic renormalization group indicate that $M_{B_c(1^-)}-M_{B_c(0^-)}=(50 \pm 17  {}^{+15}_{-12})\,\rm{MeV}$
 \cite{Penin2004}, the mass $M_{B_c^*}=6.337\,\rm{GeV}$ from the QCDSR  is satisfactory. Accordingly, we take the threshold parameters and Borel  parameters
as  $s_0=u_0=(45\pm1)\,\rm{GeV}^2$,  $M_1^2=M_2^2=(5-7)\,\rm{GeV}^2$ from the QCDSR  \cite{Wang-BC}. The uncertainties of the hadronic coupling constants
$ \delta\, G_{B_c^*B_c\Upsilon}(Q^2)$, $\delta\, G_{B_cB_c\Upsilon}(Q^2)$,   $\delta \, G_{B_c^*B_c J/\psi}(Q^2)$ and $ \delta\, G_{B_cB_c J/\psi}(Q^2)$ originate from the decay constants $f_i$ can be estimated as $ \frac{\delta f_{i}}{f_{i}}$, where $i=\Upsilon,\, J/\psi,\,B_c^*,\,B_c$. For more references on the decay constants $f_{B_c^*}$ and $f_{B_c}$, one can consult Ref.\cite{Wang2013}.

The value of the gluon condensate $\langle \frac{\alpha_s
GG}{\pi}\rangle $ has been updated from time to time, and changes
greatly, we use the recently updated value $\langle \frac{\alpha_s GG}{\pi}\rangle=(0.022 \pm
0.004)\,\rm{GeV}^4 $ \cite{gg-conden}.
For the heavy quark masses, we take the $\overline{MS}$ masses $m_{c}(m_c^2)=(1.275\pm0.025)\,\rm{GeV}$ and  $m_{b}(m_b^2)=(4.18\pm 0.03)\,\rm{GeV}$
 from the Particle Data Group \cite{PDG}, and   account for
the energy-scale dependence of  the $\overline{MS}$ masses,
\begin{eqnarray}
m_c(\mu^2)&=&m_c(m_c^2)\left[\frac{\alpha_{s}(\mu)}{\alpha_{s}(m_c)}\right]^{\frac{12}{25}} \, ,\nonumber\\
m_b(\mu^2)&=&m_b(m_b^2)\left[\frac{\alpha_{s}(\mu)}{\alpha_{s}(m_b)}\right]^{\frac{12}{23}} \, ,\nonumber\\
\alpha_s(\mu)&=&\frac{1}{b_0t}\left[1-\frac{b_1}{b_0^2}\frac{\log t}{t} +\frac{b_1^2(\log^2{t}-\log{t}-1)+b_0b_2}{b_0^4t^2}\right]\, ,
\end{eqnarray}
  where $t=\log \frac{\mu^2}{\Lambda^2}$, $b_0=\frac{33-2n_f}{12\pi}$, $b_1=\frac{153-19n_f}{24\pi^2}$, $b_2=\frac{2857-\frac{5033}{9}n_f+\frac{325}{27}n_f^2}{128\pi^3}$,  $\Lambda=213\,\rm{MeV}$, $296\,\rm{MeV}$  and  $339\,\rm{MeV}$ for the flavors  $n_f=5$, $4$ and $3$, respectively  \cite{PDG}. In this article, we take the typical energy scale $\mu=  2\,\rm{GeV}$ as in Refs.\cite{Wang-BC,Wang2013}.

In Fig.4, we plot the contributions to the hadronic coupling constants  $ G_{B_c^*B_c\Upsilon}(Q^2)$, $ G_{B_c^*B_c J/\psi}(Q^2)$,  $ G_{B_cB_c\Upsilon}(Q^2)$ and $ G_{B_cB_c J/\psi}(Q^2)$ from different terms in the operator product expansion at the value $Q^2=1\,\rm{GeV}^2$ with variations of the Borel parameters $M_1^2$ and $M_2^2$. From the figure, we can see that the  values are rather stable with variations of the  Borel parameters, Borel platforms appear. The ratios among  the  perturbative contributions, gluon condensate contributions, leading order  Coulomb-like corrections (${\mathcal{O}}(\alpha_s^{\mathcal{C}}/v_s,\alpha_s^{\mathcal{C}}/v_u)$), total Coulomb-like corrections, total contributions
 are about $(20-25)\%:(1-8)\%:50\%:(70-80)\%:1$. Although the contributions of the leading order Coulomb-like corrections are twice as large  as that of the perturbative terms, the  Coulomb-like corrections decrease quickly with increase of the orders of  $\alpha_s^{\mathcal{C}}/v_s,\alpha_s^{\mathcal{C}}/v_u$,
 \begin{eqnarray}
\frac{4\pi\alpha_s^\mathcal{C}}{3v} \frac{1}{1-\exp\left(-\frac{4\pi\alpha_s^\mathcal{C}}{3v}\right)}&=&1+\frac{2\pi\alpha_s^\mathcal{C}}{3v}+\frac{1}{12}\left(\frac{4\pi\alpha_s^\mathcal{C}}{3v}\right)^2
-\frac{1}{720}\left(\frac{4\pi\alpha_s^\mathcal{C}}{3v}\right)^4+\cdots\,   ,
\end{eqnarray}
where the $v$ denotes the $v_s$ and $v_u$,
 the operator product expansion is well convergent.

 In calculations, we observe that $0.0001\leq \exp(-\frac{s_0}{M_1^2})\leq 0.00186$ and $0.0001\leq \exp(-\frac{u_0}{M_2^2})\leq 0.00186 $, the contributions of high resonances and continuum states are greatly suppressed, the hadronic coupling constants $ G_{B_c^*B_c\Upsilon}(Q^2)$, $ G_{B_c^*B_c J/\psi}(Q^2)$, $ G_{B_cB_c\Upsilon}(Q^2)$ and $ G_{B_cB_c J/\psi}(Q^2)$ are not sensitive to the threshold parameters.   The two criteria (pole dominance and convergence of the operator product
expansion) of the QCDSR  are fully  satisfied. Furthermore, there exist Borel platforms to extract the numerical values of the hadronic coupling constants
$G_{B_c^*B_c\Upsilon}(Q^2)$, $ G_{B_c^*B_c J/\psi}(Q^2)$, $ G_{B_cB_c\Upsilon}(Q^2)$ and $ G_{B_cB_c J/\psi}(Q^2)$.

The numerical values of the hadronic coupling constants  $ G_{B_c^*B_c\Upsilon}(Q^2)$, $ G_{B_c^*B_c J/\psi}(Q^2)$, $ G_{B_cB_c\Upsilon}(Q^2)$ and $ G_{B_cB_c J/\psi}(Q^2)$ are shown explicitly in Figs.5, and fitted into the following  analytical functions by the ${ \bf \rm MINUIT}$,
 \begin{eqnarray}
 G_{B_c^*B_c\Upsilon}(Q^2)&=&A\exp\left(-BQ^2\right)\, , \nonumber\\
   G_{B_c^*B_c J/\psi}(Q^2)&=&\frac{C}{1+DQ^2+EQ^4}\exp\left( -FQ^2\right)+H\, , \nonumber\\
   G_{B_cB_c\Upsilon}(Q^2)&=&\frac{A^{\prime}}{1+B^{\prime}Q^2}\, , \nonumber\\
   G_{B_cB_c J/\psi}(Q^2)&=&\frac{C^{\prime}}{1+D^{\prime}Q^2+E^{\prime}Q^4}\exp\left( -F^{\prime}Q^2\right)+H^{\prime}\, ,
   \end{eqnarray}
   where
    \begin{eqnarray}
    A&=&3.0667\pm 0.43429\, \rm{GeV}^{-1} \, ,\nonumber\\
    B&=&0.037120\pm 0.040971\,\rm{GeV}^{-2} \, ,\nonumber\\
    C&=&6.1944\pm 11.883\, \rm{GeV}^{-1} \, ,\nonumber\\
    D&=&0.18488\pm 3.1480 \,\rm{GeV}^{-2}\, ,\nonumber\\
    E&=&0.063663\pm  1.2289\,\rm{GeV}^{-4} \, ,\nonumber\\
    F&=&0.29199\pm 2.9767 \,\rm{GeV}^{-2}\, ,\nonumber\\
    H&=&0.044515\pm   3.6666  \,\rm{GeV}^{-1}\, , \nonumber\\
    A^{\prime}&=&12.802\pm 0.68184     \, ,\nonumber\\
    B^{\prime}&=&0.0078868\pm 0.016091\,\rm{GeV}^{-2} \, ,\nonumber\\
    C^{\prime}&=&6.6854\pm 9.7404 \, ,\nonumber\\
    D^{\prime}&=&0.31276\pm 1.9985 \,\rm{GeV}^{-2}\, ,\nonumber\\
    E^{\prime}&=&0.086261\pm  1.0481\,\rm{GeV}^{-4} \, ,\nonumber\\
    F^{\prime}&=&0.14764\pm 1.7194 \,\rm{GeV}^{-2}\, ,\nonumber\\
    H^{\prime}&=&0.20721\pm   2.2281  \, .
     \end{eqnarray}
 Although  the  uncertainties of the parameters in the $G_{B_c^*B_c J/\psi}(Q^2)$, $G_{B_cB_c J/\psi}(Q^2)$ are very large, the central values of the fitted functions
 $ G_{B_c^*B_c\Upsilon}(Q^2)$, $G_{B_c^*B_c J/\psi}(Q^2)$, $ G_{B_cB_c\Upsilon}(Q^2)$ and $G_{B_cB_c J/\psi}(Q^2)$  coincide with  the central values from the
 QCDSR.

 From the numerical values of the $ G_{B_c^*B_c\Upsilon}(Q^2)$, $ G_{B_c^*B_c J/\psi}(Q^2)$, $ G_{B_cB_c\Upsilon}(Q^2)$ and $ G_{B_cB_c J/\psi}(Q^2)$ at $Q^2=1\,\rm{GeV}^2$,
\begin{eqnarray}
 G_{B_c^*B_c\Upsilon}(Q^2=1\,{\rm GeV^2})&=&3.0\pm 0.6 \,\rm{GeV}^{-1}\, , \nonumber\\
   G_{B_c^*B_c J/\psi}(Q^2=1\,{\rm GeV^2})&=&3.7\pm 0.8 \,\rm{GeV}^{-1}\, , \nonumber\\
   G_{B_cB_c\Upsilon}(Q^2=1\,{\rm GeV^2})&=&12.8\pm2.3 \, , \nonumber\\
   G_{B_cB_c J/\psi}(Q^2=1\,{\rm GeV^2})&=&4.3 \pm 0.9\, ,
   \end{eqnarray}
we can obtain the ratios,
\begin{eqnarray}
\frac{ G_{B_cB_c\Upsilon}(Q^2=1\,{\rm GeV^2})}{G_{B_c^*B_c\Upsilon}(Q^2=1\,{\rm GeV^2})}&=&4.3\pm1.2\,{\rm GeV} \approx m_b(m_b^2) \, ,\nonumber \\
\frac{ G_{B_cB_cJ/\psi}(Q^2=1\,{\rm GeV^2})}{G_{B_c^*B_cJ\psi}(Q^2=1\,{\rm GeV^2})}&=&1.2\pm0.4\,{\rm GeV}\approx  m_c(m_c^2) \, ,
\end{eqnarray}
if the uncertainties are neglected. The ratio $\frac{ G_{B_cB_c\Upsilon}(Q^2)}{G_{B_c^*B_c\Upsilon}(Q^2)}$ increases slowly with increase of the $Q^2$, while the ratio $\frac{ G_{B_cB_cJ/\psi}(Q^2)}{G_{B_c^*B_cJ\psi}(Q^2)}$ increases quickly with increase of the $Q^2$ at the range $Q^2=(1-6)\,\rm{GeV}^2$.

 In Fig.6, we extrapolate  the hadronic coupling constants $ G_{B_c^*B_c\Upsilon}(Q^2)$, $G_{B_c^*B_c J/\psi}(Q^2)$, $ G_{B_cB_c\Upsilon}(Q^2)$ and $G_{B_cB_c J/\psi}(Q^2)$  into the deep time-like regions analytically. From the figure, we can see that the $ G_{B_c^*B_c\Upsilon}(Q^2)$ and $ G_{B_cB_c\Upsilon}(Q^2)$ increase monotonously with  increase of the squared momentum $q^2=-Q^2$, the $G_{B_c^*B_c J/\psi}(Q^2)$ also increases steadily with increase of the squared momentum $q^2=-Q^2$ and develops a shoulder at about $q^2=3\,\rm{GeV}^2$, while the $G_{B_cB_c J/\psi}(q^2)$ develops  a broad peak at about $q^2=2.4\,\rm{GeV}^2$. The    fitted functions  $G_{B_c^*B_c J/\psi}(q^2)$ and $G_{B_cB_c J/\psi}(q^2)$ in the time-like region  $q^2=-Q^2>0$ can be reexpressed in the following forms,
 \begin{eqnarray}
   G_{B_c^*B_c J/\psi}(q^2)&=&\frac{C\exp\left( Fq^2\right)}{1-Dq^2+Eq^4}+H=\frac{C\exp\left( Fq^2\right)}{\left(1-\frac{D}{2}q^2\right)^2+\left(E-\frac{D^2}{4}\right)q^4}+H\, , \nonumber\\
      G_{B_cB_c J/\psi}(q^2)&=&\frac{C^{\prime}\exp\left( F^{\prime}q^2\right)}{1-D^{\prime}q^2+E^{\prime}q^4}+H^{\prime}=\frac{C^{\prime}\exp\left( F^{\prime}q^2\right)}{\left(1-\frac{D^{\prime}}{2}q^2\right)^2+\left(E^{\prime}-\frac{D^{\prime2}}{4}\right)q^4}
      +H^{\prime}\, ,
   \end{eqnarray}
with $E-\frac{D^2}{4}>0$ and $E^{\prime}-\frac{D^{\prime2}}{4}>0$ for the central values of the parameters. There maybe appear peaks  at the neighborhood of the values
$q^2=\frac{1}{D}$, $\frac{2}{D}$,  $\frac{1}{D^{\prime}}$ and $\frac{2}{D^{\prime}}$.

The extrapolation to deep time-like regions is highly mode-dependent and leads to  systematic  uncertainties for  the hadronic coupling constants.
In order to minimize the systematic uncertainties, we can study the vertices
  simultaneously   by  putting the $B_c^*$, $B_c$, $\Upsilon$ (or $J/\psi$) off-shell sequentially, then fit the hadronic coupling constants to suitable analytical functions and extrapolate them to the physical regions by
 requiring the on-shell values of the   hadronic coupling constants coincide \cite{3PTQCDSR-Rev}.
We postpone  the tedious calculations to our next work.

 Finally, we obtain the one-shell  values of the  $ G_{B_c^*B_c\Upsilon}(q^2)$,  $G_{B_cB_c\Upsilon}(q^2)$, $ G_{B_c^*B_c J/\psi}(q^2)$ and $ G_{B_cB_c J/\psi}(q^2)$ from the fitted functions,
\begin{eqnarray}
 G_{B_c^*B_c\Upsilon}(q^2=M_{\Upsilon}^2)&=&85\,\rm{GeV}^{-1}\, , \nonumber\\
   G_{B_c^*B_c J/\psi}(q^2=M^2_{J/\psi})&=&20\,\rm{GeV}^{-1}\, , \nonumber\\
   G_{B_cB_c\Upsilon}(q^2=M_{\Upsilon}^2)&=&44 \, , \nonumber\\
   G_{B_cB_c J/\psi}(q^2=M^2_{J/\psi})&=&5 \, ,
   \end{eqnarray}
where we retain the central values only, as  the uncertainties are too large to make sense. The uncertainties originate from the uncertainties $\delta B$, $\delta C$, $\delta D$, $\delta E$, $\delta F$, $\delta B^{\prime}$, $\delta C^{\prime}$, $\delta D^{\prime}$, $\delta E^{\prime}$ and $\delta F^{\prime}$
 are greatly amplified in the deep time-like regions, and much larger than the central values,   while the uncertainties originate from the uncertainties $\delta A$ and $\delta A^{\prime}$ are moderate.
For example, the uncertainties $\delta A=\pm 0.43429\,\rm{GeV}^{-1}$, $\delta B=\pm 0.040971\,\rm{GeV}^{-2}$, $\delta A^{\prime}=\pm 0.68184$ and $\delta B^{\prime}=\pm0.016091\,\rm{GeV}^{-2}$ lead to $\delta G_{B_c^*B_c\Upsilon}(q^2=M_{\Upsilon}^2)=\pm 12\,\rm{GeV}^{-1}$,   $\delta G_{B_c^*B_c\Upsilon}(q^2=M_{\Upsilon}^2)=\pm 312\,\rm{GeV}^{-1}$, $\delta G_{B_cB_c\Upsilon}(q^2=M_{\Upsilon}^2)=\pm 2$ and   $\delta G_{B_cB_c\Upsilon}(q^2=M_{\Upsilon}^2)=\pm 213$, respectively. It is obvious that the uncertainties  $\delta G_{B_c^*B_c\Upsilon}(q^2=M_{\Upsilon}^2)=\pm 312\,\rm{GeV}^{-1}$  and   $\delta G_{B_cB_c\Upsilon}(q^2=M_{\Upsilon}^2)=\pm 213$ are too large to make sense.
On the other hand, although the uncertainties $\delta H$ and $\delta H^{\prime}$ do not vary with the $q^2$, they are about ten times as large as  the corresponding central values. So we only retain the central values, which are more reasonable than the uncertainties.
We can take those hadronic coupling constants as basic input parameters  to  study
final-state interactions in the heavy quarkonium decays, or calculate the   absorption cross sections at the hadronic level   to understand  the  heavy quarkonium  absorptions in hadronic matter.

\begin{figure}
 \centering
 \includegraphics[totalheight=6cm,width=7cm]{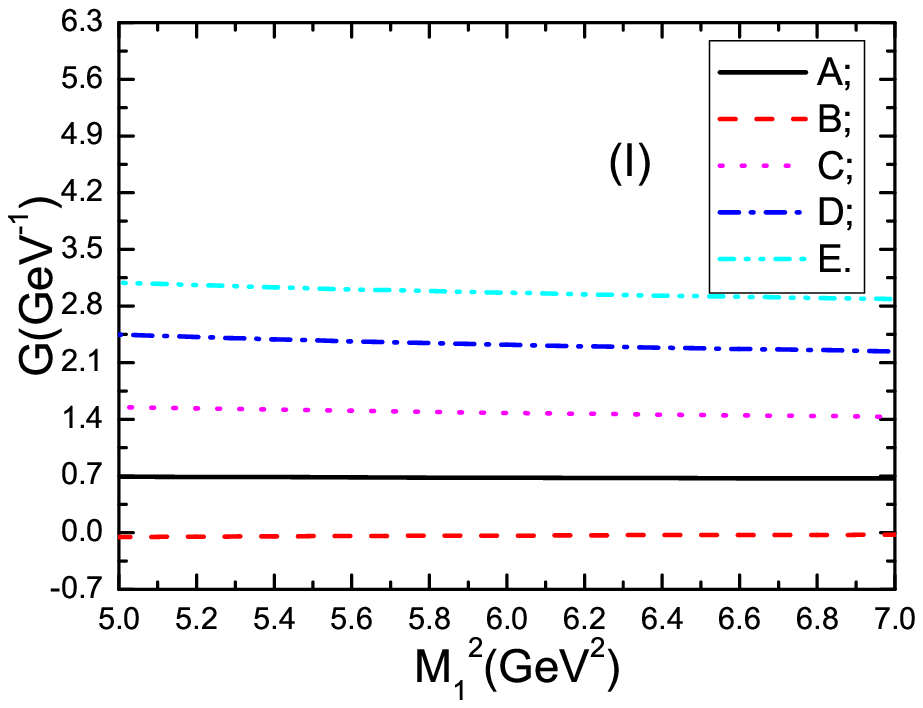}
 \includegraphics[totalheight=6cm,width=7cm]{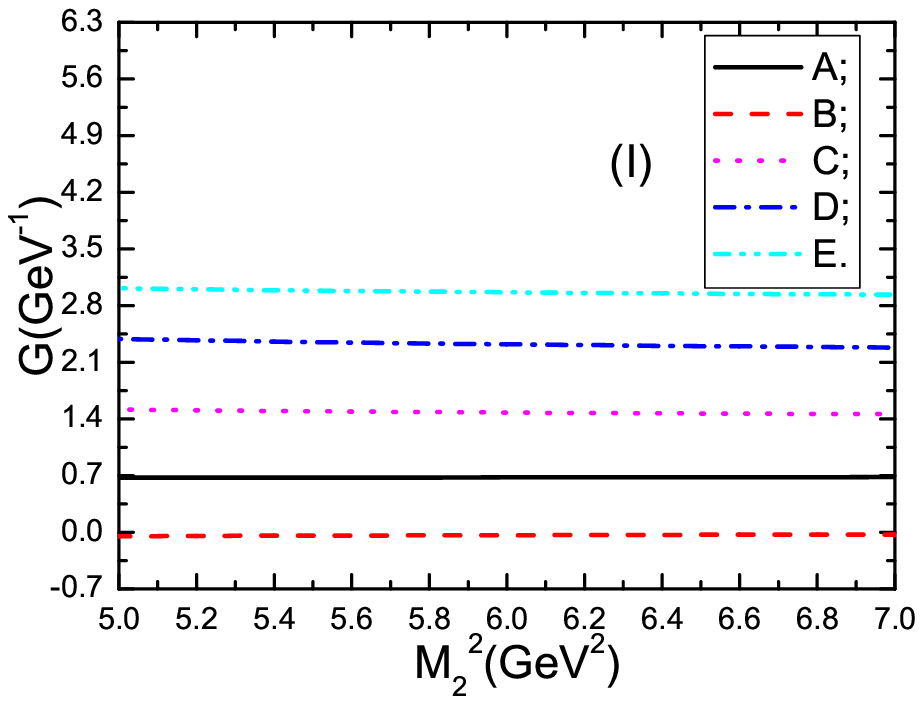}
 \includegraphics[totalheight=6cm,width=7cm]{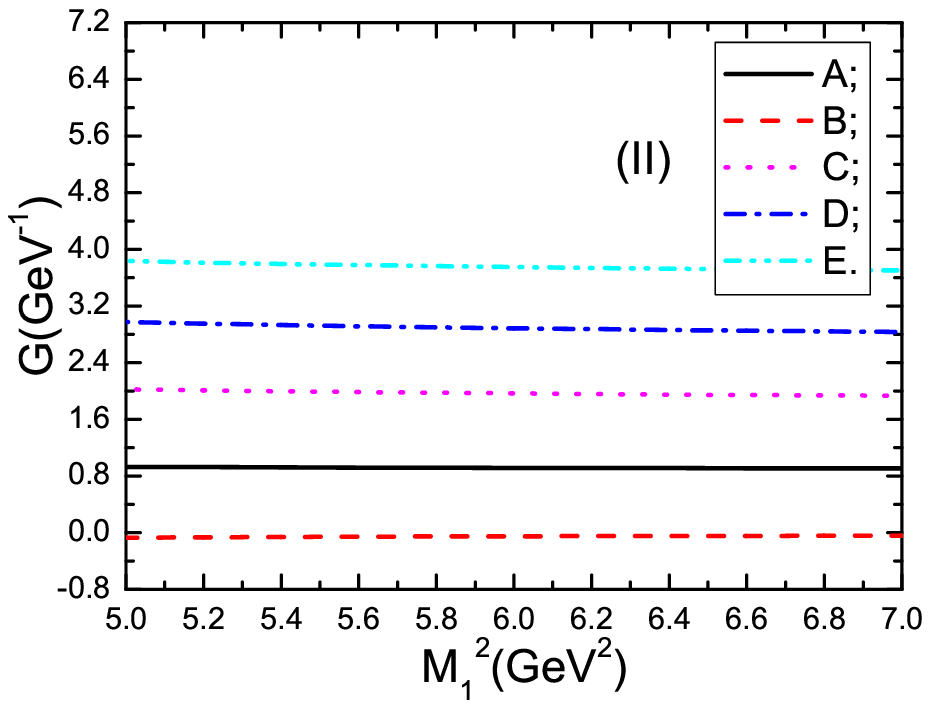}
 \includegraphics[totalheight=6cm,width=7cm]{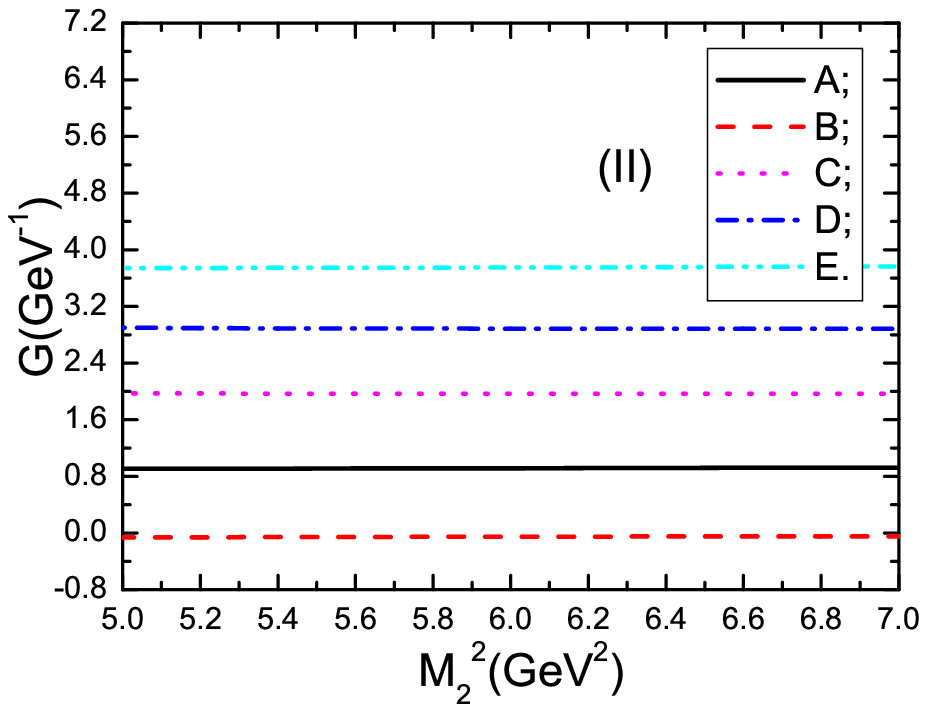}
 \includegraphics[totalheight=6cm,width=7cm]{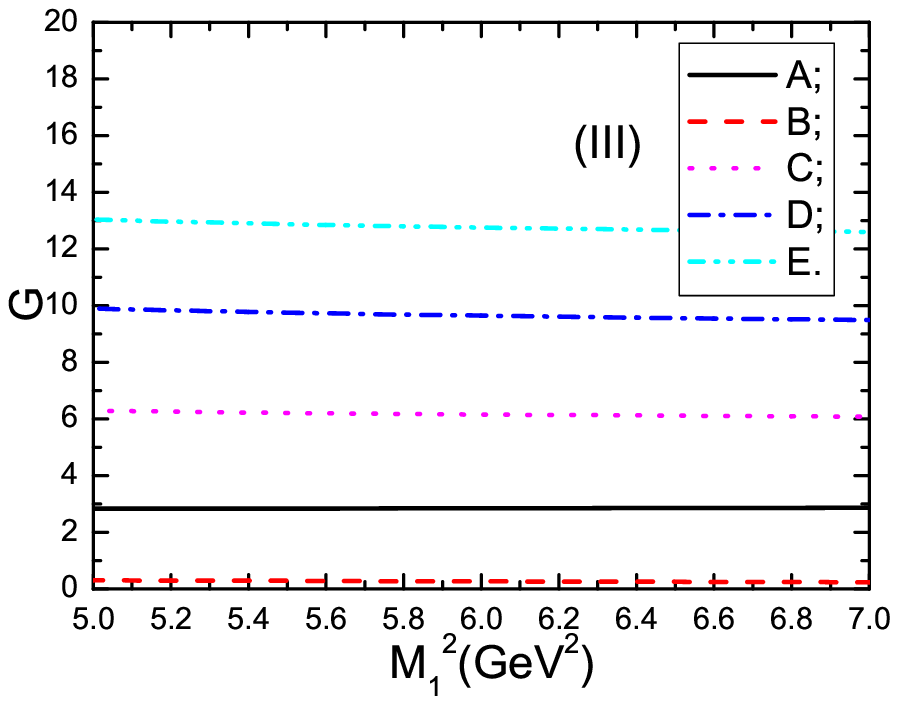}
 \includegraphics[totalheight=6cm,width=7cm]{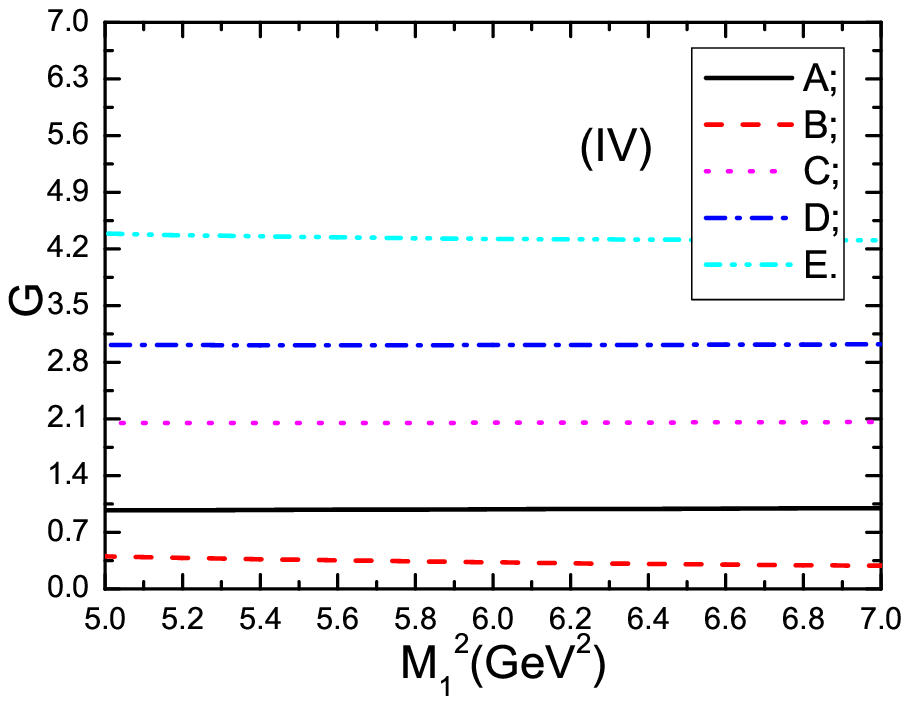}
        \caption{ The hadronic coupling constants $G_{B_c^*B_c\Upsilon}(Q^2)$ (I), $G_{B_c^*B_cJ/\psi}(Q^2)$ (II), $G_{B_cB_c\Upsilon}(Q^2)$ (III) and $G_{B_cB_cJ/\psi}(Q^2)$ (IV)   with variations of the Borel parameters $M_1^2$ or $M_2^2$ at the value $Q^2=1\,\rm{GeV}^2$. The $A$, $B$, $C$, $D$ and $E$ denote the  perturbative contributions, gluon condensate contributions, leading order  Coulomb-like corrections (${\mathcal{O}}(\alpha_s^{\mathcal{C}}/v_s,\alpha_s^{\mathcal{C}}/v_u)$), total Coulomb-like corrections and total contributions, respectively. The values of un-plotted parameters are  tacitly taken as $M_1^2=6\,\rm{GeV}^2$ or $M_2^2=6\,\rm{GeV}^2$.   }
\end{figure}

\begin{figure}
 \centering
 \includegraphics[totalheight=6cm,width=7cm]{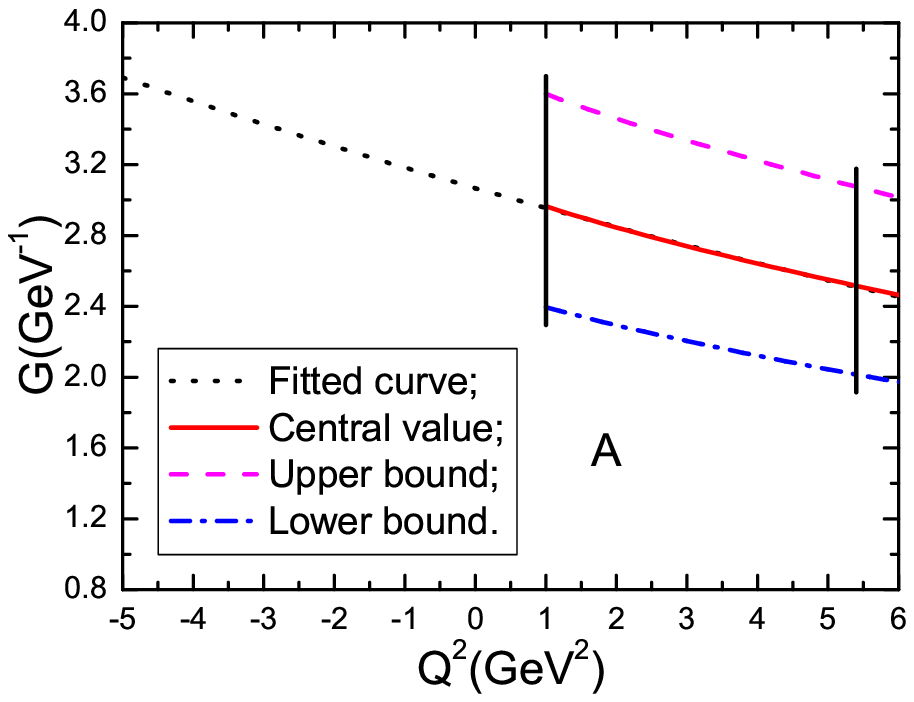}
\includegraphics[totalheight=6cm,width=7cm]{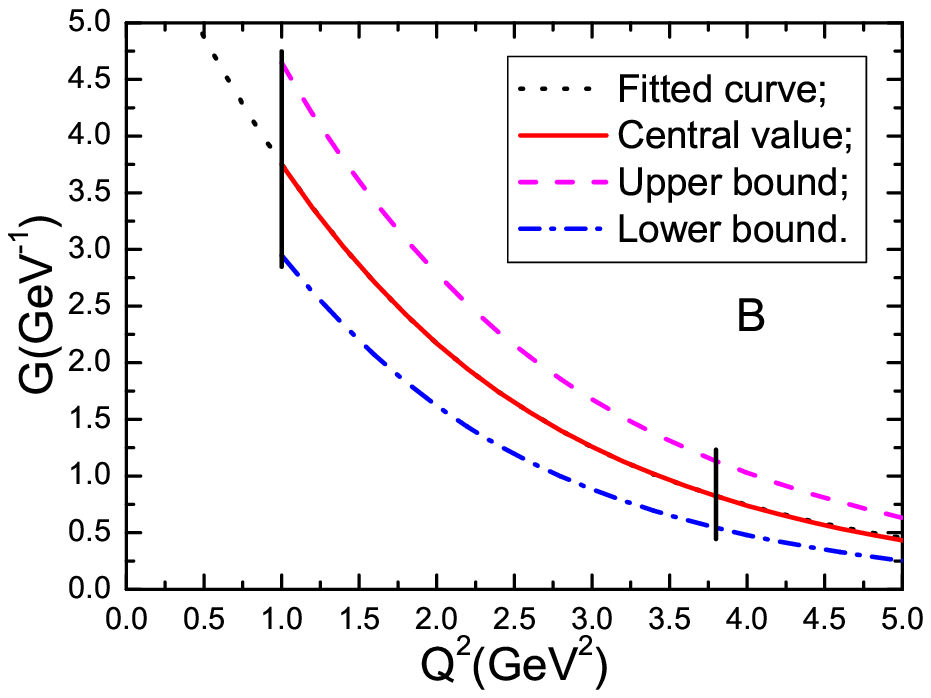}
\includegraphics[totalheight=6cm,width=7cm]{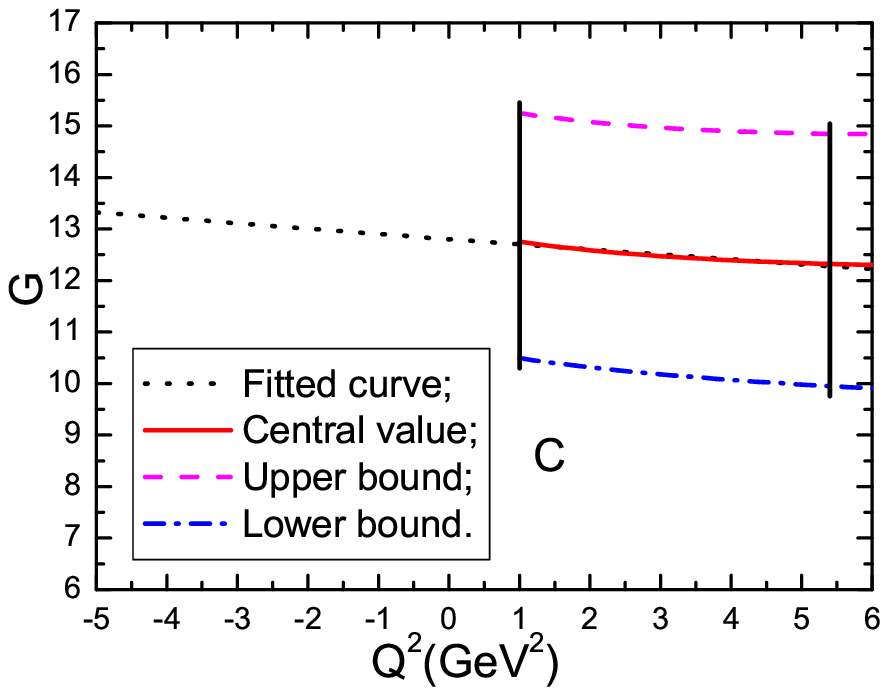}
\includegraphics[totalheight=6cm,width=7cm]{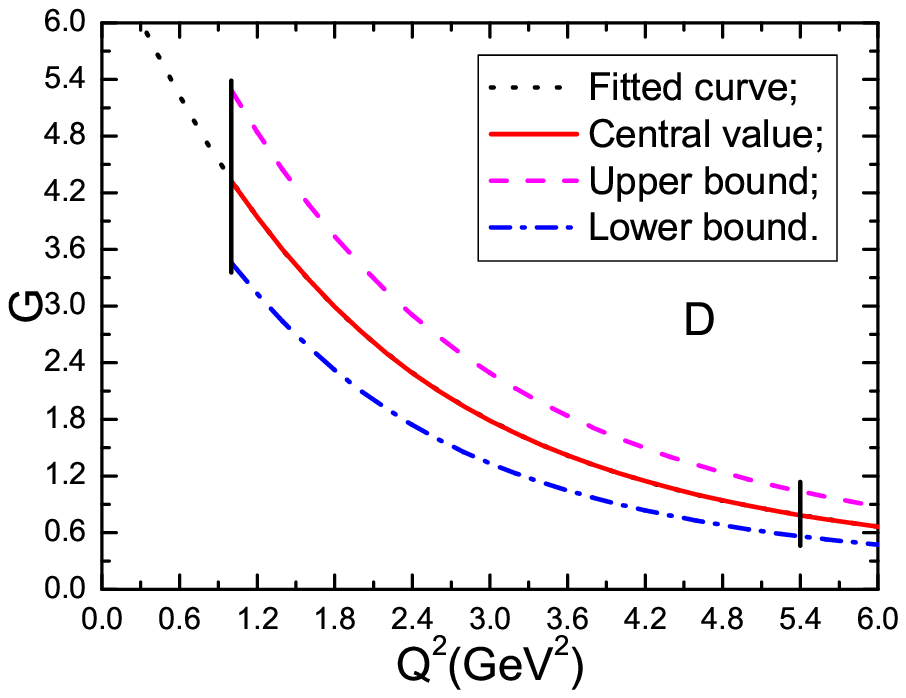}
        \caption{ The hadronic coupling constants $G_{B_c^*B_c\Upsilon}(Q^2)$ ($A$), $G_{B_c^*B_cJ/\psi}(Q^2)$ ($B$), $G_{B_cB_c\Upsilon}(Q^2)$ ($C$) and $G_{B_cB_cJ/\psi}(Q^2)$ ($D$) with variations of the $Q^2=-q^2$, where the fitted curve denotes the central values of the fitted functions.  The data between the two  perpendicular lines are used to fit the parameters of the hadronic coupling constants. }
\end{figure}
\begin{figure}
 \centering
 \includegraphics[totalheight=6cm,width=7cm]{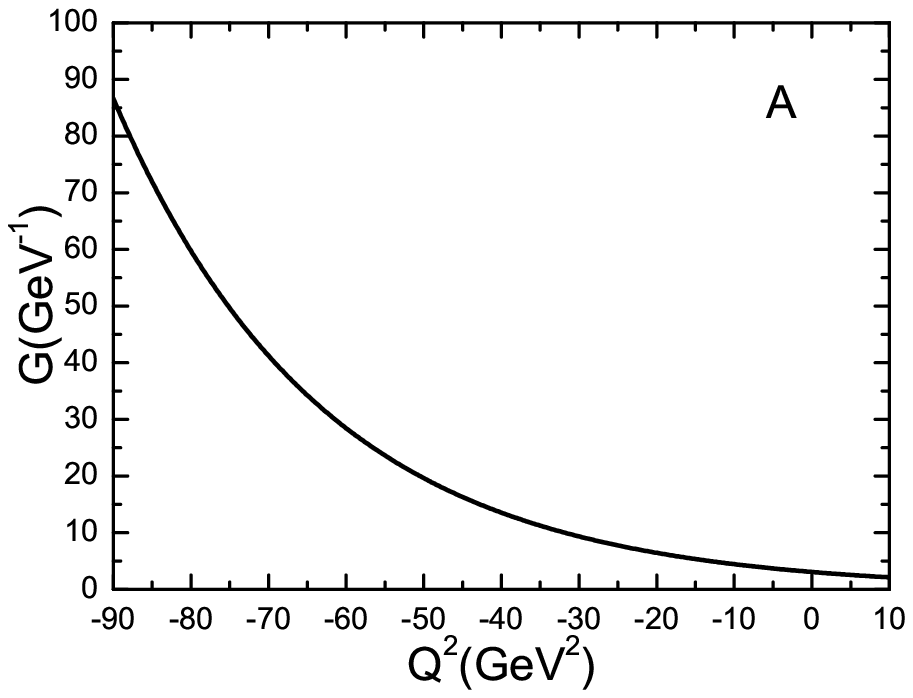}
\includegraphics[totalheight=6cm,width=7cm]{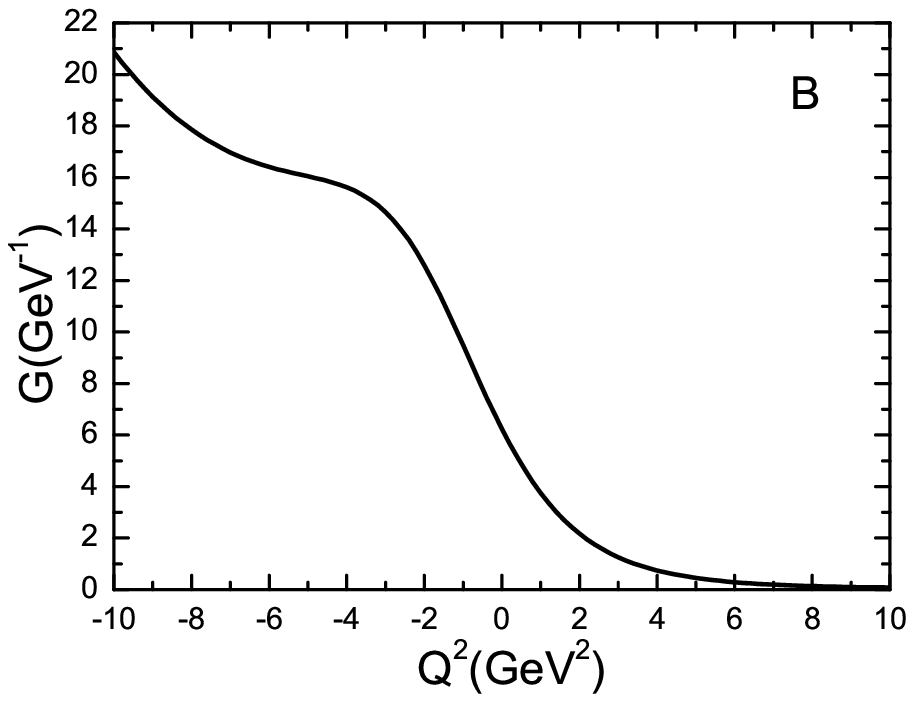}
 \includegraphics[totalheight=6cm,width=7cm]{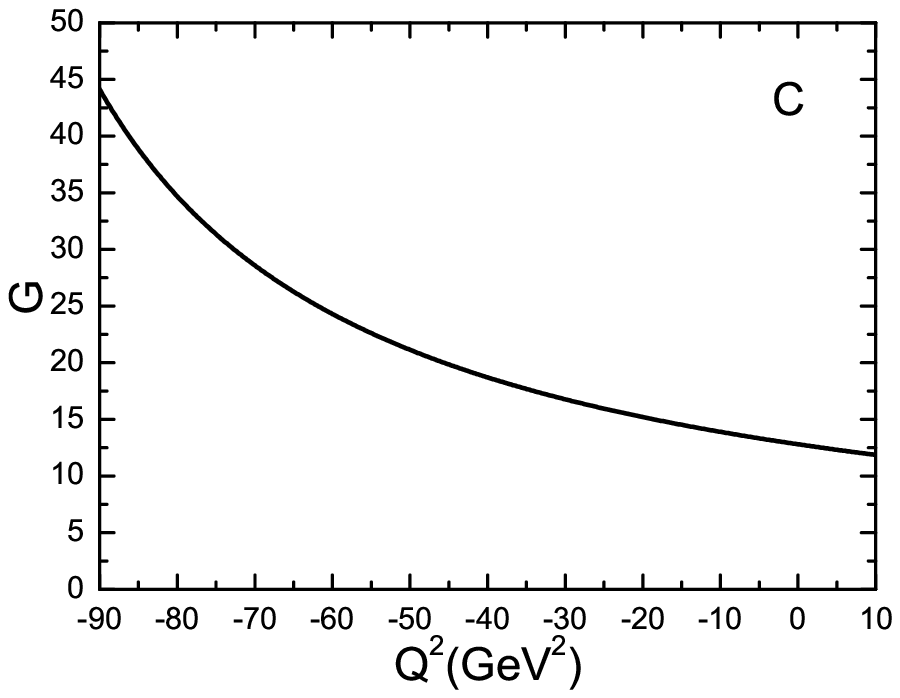}
 \includegraphics[totalheight=6cm,width=7cm]{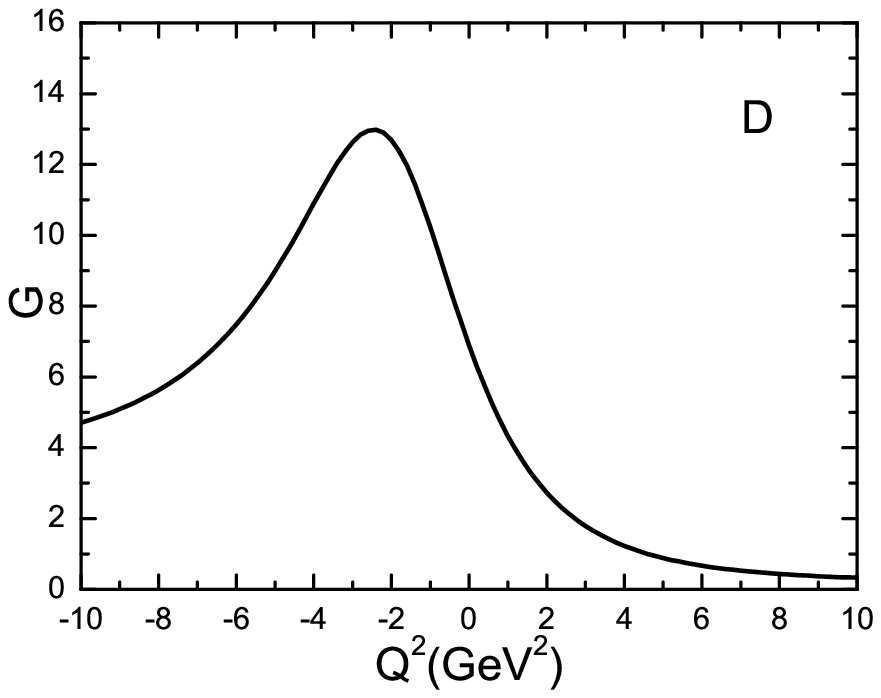}
        \caption{ The central values of the hadronic coupling constants $G_{B_c^*B_c\Upsilon}(Q^2)$ ($A$), $G_{B_c^*B_cJ/\psi}(Q^2)$ ($B$), $G_{B_cB_c\Upsilon}(Q^2)$ ($C$) and $G_{B_cB_cJ/\psi}(Q^2)$ ($D$)
         extrapolated  into the time-like regions. }
\end{figure}

\section{Conclusion}
In this article, we study  the  momentum dependence of the hadronic coupling constants $G_{B_c^* B_c \Upsilon}$, $G_{B_c^* B_c J/\psi}$, $G_{B_c B_c \Upsilon}$ and $G_{B_c B_c J/\psi}$
with the off-shell $\Upsilon$ and $J/\psi$  using the three-point  QCDSR. Then we fit the  hadronic coupling constants
$G_{B_c^*B_c\Upsilon}(Q^2)$, $ G_{B_c^*B_c J/\psi}(Q^2)$, $G_{B_cB_c\Upsilon}(Q^2)$ and $ G_{B_cB_c J/\psi}(Q^2)$ into   analytical functions, extrapolate  them into the deep time-like regions, and obtain the one-shell  values  $ G_{B_c^*B_c\Upsilon}(Q^2=-M_{\Upsilon}^2)$, $ G_{B_c^*B_c J/\psi}(Q^2=-M_{J/\psi}^2)$,
$ G_{B_cB_c\Upsilon}(Q^2=-M_{\Upsilon}^2)$ and $ G_{B_cB_c J/\psi}(Q^2=-M_{J/\psi}^2)$ for the first time, no other theoretical work on this subject exist.
 The hadronic coupling constants can be taken as basic input parameters in studying  the heavy quarkonium absorptions in hadronic matter and
final-state interactions in the heavy quarkonium  hadronic decays.

\section*{Acknowledgements}
This  work is supported by National Natural Science Foundation,
Grant Number 11375063,  and the Fundamental Research Funds for the
Central Universities.

\section*{Appendix}
The explicit expressions of the $I_0^{ijn}$, $K_0^{ijn}$, $N_0^{ijn}$, $I_{10}^{ijn}$, $I_{01}^{ijn}$, $K_{10}^{ijn}$, $K_{01}^{ijn}$, $N_{10}^{ijn}$, $N_{01}^{ijn}$,
\begin{eqnarray}
iI_0^{ijn}&=&B_{-p^2\rightarrow M_1^2}B_{-p^{\prime2}\rightarrow M_2^2} \overline{I}_{ijn} \nonumber\\
&=&\frac{(-1)^{i+j+n}i\pi^2}{\Gamma(i)\Gamma(j)\Gamma(n)\left(M_2^2\right)^i\left(M_1^2\right)^j \left(M^2\right)^{n-2}}\int_0^1 d\lambda \frac{\lambda^{1-i-j}}{(1-\lambda)^{n-1}} \nonumber\\
&&\exp\left\{ -\frac{(1-\lambda)Q^2}{\lambda\left(M_1^2+M_2^2\right)}-\frac{m_b^2}{\lambda M^2}-\frac{m_c^2}{(1-\lambda)M^2}\right\} \, ,
\end{eqnarray}

\begin{eqnarray}
iK_0^{ijn}&=&B_{-p^2\rightarrow M_1^2}B_{-p^{\prime2}\rightarrow M_2^2} \overline{K}_{ijn} \nonumber\\
&=&\frac{d}{dt}\frac{(-1)^{i+j+n}i\pi^2}{\Gamma(i)\Gamma(j)\Gamma(n)\left(\overline{M}_2^2\right)^i\left(\overline{M}_1^2\right)^{j-1} \left(\overline{M}^2\right)^{n-2}}\int_0^1 d\lambda \frac{\lambda^{1-i-j}}{(1-\lambda)^{n-1}} \nonumber\\
&&\exp\left\{ -\frac{(1-\lambda)Q^2}{\lambda\left(\overline{M}_1^2+\overline{M}_2^2\right)}-\frac{m_b^2}{\lambda \overline{M}^2}-\frac{m_c^2}{(1-\lambda)\overline{M}^2}\right\}|_{t=1,r=1} \, ,
\end{eqnarray}

\begin{eqnarray}
iN_0^{ijn}&=&B_{-p^2\rightarrow M_1^2}B_{-p^{\prime2}\rightarrow M_2^2} \overline{N}_{ijn} \nonumber\\
&=&\frac{d}{dr}\frac{(-1)^{i+j+n}i\pi^2}{\Gamma(i)\Gamma(j)\Gamma(n)\left(\overline{M}_2^2\right)^{i-1}\left(\overline{M}_1^2\right)^{j} \left(\overline{M}^2\right)^{n-2}}\int_0^1 d\lambda \frac{\lambda^{1-i-j}}{(1-\lambda)^{n-1}} \nonumber\\
&&\exp\left\{ -\frac{(1-\lambda)Q^2}{\lambda\left(\overline{M}_1^2+\overline{M}_2^2\right)}-\frac{m_b^2}{\lambda \overline{M}^2}-\frac{m_c^2}{(1-\lambda)\overline{M}^2}\right\}|_{t=1,r=1} \, ,
\end{eqnarray}

\begin{eqnarray}
iI^\mu_{ijn}&=&B_{-p^2\rightarrow M_1^2}B_{-p^{\prime 2}\rightarrow M_2^2} \overline{I}^\mu_{ijn} \nonumber\\
&=&\frac{(-1)^{i+j+n+1}i\pi^2}{\Gamma(i)\Gamma(j)\Gamma(n)\left(M_2^2\right)^i\left(M_1^2\right)^{j+1} \left(M^2\right)^{n-3}}\int_0^1 d\lambda \frac{\lambda^{1-i-j}}{(1-\lambda)^{n-2}} \nonumber\\
&&\exp\left\{ -\frac{(1-\lambda)Q^2}{\lambda\left(M_1^2+M_2^2\right)}-\frac{m_b^2}{\lambda M^2}-\frac{m_c^2}{(1-\lambda)M^2}\right\} (p-p^{\prime})^\mu \nonumber\\
&&+\frac{(-1)^{i+j+n}i\pi^2}{\Gamma(i)\Gamma(j)\Gamma(n)\left(M_2^2\right)^i\left(M_1^2\right)^j \left(M^2\right)^{n-2}}\int_0^1 d\lambda \frac{\lambda^{2-i-j}}{(1-\lambda)^{n-1}} \nonumber\\
&&\exp\left\{ -\frac{(1-\lambda)Q^2}{\lambda\left(M_1^2+M_2^2\right)}-\frac{m_b^2}{\lambda M^2}-\frac{m_c^2}{(1-\lambda)M^2}\right\} p^{\prime \mu} \nonumber\\
&=&iI_{10}^{ijn}(p-p^{\prime})^\mu+iI_{01}^{ijn}p^{\prime\mu}\, ,
\end{eqnarray}

\begin{eqnarray}
iK^\mu_{ijn}&=&B_{-p^2\rightarrow M_1^2}B_{-p^{\prime 2}\rightarrow M_2^2} \overline{K}^\mu_{ijn} \nonumber\\
&=&\frac{d}{dt}\frac{(-1)^{i+j+n+1}i\pi^2}{\Gamma(i)\Gamma(j)\Gamma(n)\left(\overline{M}_2^2\right)^i\left(\overline{M}_1^2\right)^{j} \left(\overline{M}^2\right)^{n-3}}\int_0^1 d\lambda \frac{\lambda^{1-i-j}}{(1-\lambda)^{n-2}} \nonumber\\
&&\exp\left\{ -\frac{(1-\lambda)Q^2}{\lambda\left(\overline{M}_1^2+\overline{M}_2^2\right)}-\frac{m_b^2}{\lambda \overline{M}^2}-\frac{m_c^2}{(1-\lambda)\overline{M}^2}\right\} |_{t=1,r=1}(p-p^{\prime})^\mu \nonumber\\
&&+\frac{d}{dt}\frac{(-1)^{i+j+n}i\pi^2}{\Gamma(i)\Gamma(j)\Gamma(n)\left(\overline{M}_2^2\right)^i\left(\overline{M}_1^2\right)^{j-1} \left(\overline{M}^2\right)^{n-2}}\int_0^1 d\lambda \frac{\lambda^{2-i-j}}{(1-\lambda)^{n-1}} \nonumber\\
&&\exp\left\{ -\frac{(1-\lambda)Q^2}{\lambda\left(\overline{M}_1^2+\overline{M}_2^2\right)}-\frac{m_b^2}{\lambda \overline{M}^2}-\frac{m_c^2}{(1-\lambda)\overline{M}^2}\right\}|_{t=1,r=1} p^{\prime \mu} \nonumber\\
&=&iK_{10}^{ijn}(p-p^{\prime})^\mu+iK_{01}^{ijn}p^{\prime\mu}\, ,
\end{eqnarray}

\begin{eqnarray}
iN^\mu_{ijn}&=&B_{-p^2\rightarrow M_1^2}B_{-p^{\prime 2}\rightarrow M_2^2} \overline{N}^\mu_{ijn} \nonumber\\
&=&\frac{d}{dr}\frac{(-1)^{i+j+n+1}i\pi^2}{\Gamma(i)\Gamma(j)\Gamma(n)\left(\overline{M}_2^2\right)^{i-1}\left(\overline{M}_1^2\right)^{j+1} \left(\overline{M}^2\right)^{n-3}}\int_0^1 d\lambda \frac{\lambda^{1-i-j}}{(1-\lambda)^{n-2}} \nonumber\\
&&\exp\left\{ -\frac{(1-\lambda)Q^2}{\lambda\left(\overline{M}_1^2+\overline{M}_2^2\right)}-\frac{m_b^2}{\lambda \overline{M}^2}-\frac{m_c^2}{(1-\lambda)\overline{M}^2}\right\} |_{t=1,r=1}(p-p^{\prime})^\mu \nonumber\\
&&+\frac{d}{dr}\frac{(-1)^{i+j+n}i\pi^2}{\Gamma(i)\Gamma(j)\Gamma(n)\left(\overline{M}_2^2\right)^{i-1}\left(\overline{M}_1^2\right)^{j} \left(\overline{M}^2 \right)^{n-2}}\int_0^1 d\lambda \frac{\lambda^{2-i-j}}{(1-\lambda)^{n-1}} \nonumber\\
&&\exp\left\{ -\frac{(1-\lambda)Q^2}{\lambda\left(\overline{M}_1^2+\overline{M}_2^2\right)}-\frac{m_b^2}{\lambda \overline{M}^2}-\frac{m_c^2}{(1-\lambda)\overline{M}^2}\right\}|_{t=1,r=1} p^{\prime \mu} \nonumber\\
&=&iN_{10}^{ijn}(p-p^{\prime})^\mu+iN_{01}^{ijn}p^{\prime\mu}\, ,
\end{eqnarray}
where
\begin{eqnarray}
M^2&=&\frac{M_1^2M_2^2}{M_1^2+M_2^2} \, , \nonumber\\
\overline{M}^2&=&\frac{\overline{M}_1^2\overline{M}_2^2}{\overline{M}_1^2+\overline{M}_2^2} \, , \nonumber\\
\overline{M}_1^2&=&t M_1^2\, , \nonumber\\
\overline{M}_2^2&=&r M_2^2 \, ,
\end{eqnarray}
  and the $B_{-p^2\rightarrow M_1^2}B_{-p^{\prime 2}\rightarrow M_2^2}$ denotes the double Borel transform.


\begin{thebibliography}{99}


\bibitem{Matsui86} T. Matsui and H. Satz,  Phys. Lett. {\bf B178} (1986) 416.


\bibitem{QGP-rev} R. Vogt, Phys. Rept. {\bf 310} (1999) 197;
R. Rapp, D. Blaschke and P. Crochet, Prog. Part. Nucl. Phys. {\bf 65} (2010) 209.


\bibitem{Recombine-cc} P. Braun-Munzinger and J. Stachel, Phys. Lett. {\bf B490} (2000) 196;
R. L. Thews, M. Schroedter and J. Rafelski, Phys. Rev. {\bf C63} (2001) 054905;
L. Grandchamp and R. Rapp, Phys. Lett. {\bf B523} (2001) 60;
L. Grandchamp and R. Rapp, Nucl. Phys. {\bf A709} (2002) 415;
  A. Capella, L. Bravina, E. G. Ferreiro, A. B. Kaidalov, K. Tywoniuk and E. Zabrodin, Eur. Phys. J. {\bf C58} (2008) 437.



\bibitem{CNM} E. G. Ferreiro, F. Fleuret, J. P. Lansberg and A. Rakotozafindrabe, Phys. Lett. {\bf B680} (2009) 50;
R. Vogt, Phys. Rev. {\bf C81} (2010) 044903;
A. Rakotozafindrabe, E. G. Ferreiro, F. Fleuret, J. P. Lansberg and N. Matagne, Nucl. Phys. {\bf A855} (2011) 327.





\bibitem{meson-lagran}
S. G. Matinyan and B. Muller, Phys. Rev. {\bf C58} (1998) 2994;
K. L. Haglin, Phys. Rev. {\bf C61} (2000) 031902;
Z. W. Lin and C. M. Ko, Phys. Rev. {\bf C62} (2000) 034903;
A. Sibirtsev, K. Tsushima and A. W. Thomas, Phys. Rev. {\bf C63} (2001) 044906;
Z. W. Lin and C. M. Ko, Phys. Lett. {\bf B503} (2001) 104.


\bibitem{HeavyQ}  R. Casalbuoni, A. Deandrea, N. Di Bartolomeo, R. Gatto, F. Feruglio and G. Nardulli, Phys. Rept. {\bf 281} (1997) 145;
X. Liu, B. Zhang and S. L. Zhu, Phys. Lett. {\bf B645} (2007) 185;
C. Meng and K. T. Chao, Phys. Rev. {\bf D78} (2008) 074001;
F. K. Guo, C. Hanhart, G. Li, U. G. Meissner and Q. Zhao, Phys. Rev. {\bf D83} (2011) 034013.





\bibitem{3PTQCDSR}   F. S. Navarra, M. Nielsen, M. E. Bracco, M. Chiapparini and C. L. Schat, Phys. Lett. {\bf B489} (2000) 319;
M. E. Bracco, M. Chiapparini, A. Lozea, F. S. Navarra and M. Nielsen, Phys. Lett. {\bf B521} (2001) 1;
F. S. Navarra, M. Nielsen and M. E. Bracco, Phys. Rev. {\bf D65} (2002) 037502;
R. D. Matheus, F.S. Navarra, M. Nielsen and R. Rodrigues da Silva, Phys. Lett. {\bf B541} (2002) 265;
 R. Rodrigues da Silva, R. D. Matheus, F. S. Navarra and M. Nielsen, Braz. J. Phys. {\bf 34} (2004) 236;
 M. E. Bracco, M. Chiapparini, F. S. Navarra and M. Nielsen, Phys. Lett. {\bf B605} (2005) 326;
  M. E. Bracco, A. Cerqueira,   M. Chiapparini, A. Lozea  and M. Nielsen, Phys. Lett. {\bf B641} (2006) 286;
  M. E. Bracco, M. Chiapparini, F. S. Navarra and M. Nielsen, Phys. Lett. {\bf B659} (2008) 559;
  M. E. Bracco and M. Nielsen, Phys. Rev. {\bf D82} (2010) 034012;
   B. O. Rodrigues, M. E. Bracco, M. Nielsen and F. S. Navarra, Nucl. Phys. {\bf A852} (2011) 127;
    K. Azizi and H. Sundu,  J. Phys. {\bf G38} (2011) 045005;
H. Sundu, J.Y. Sungu, S. Sahin, N. Yinelek and K. Azizi, Phys. Rev. {\bf D83} (2011) 114009;
 A. Cerqueira, Jr, B. O. Rodrigues and M. E. Bracco, Nucl. Phys. {\bf A874} (2012) 130;
  C. Y. Cui, Y. L. Liu and M. Q. Huang, Phys. Lett. {\bf B707} (2012) 129;
C. Y. Cui, Y. L. Liu and M. Q. Huang, Phys. Lett. {\bf B711} (2012) 317.

\bibitem{3PTQCDSR-Rev}
M. E. Bracco, M. Chiapparini, F. S. Navarra and M. Nielsen, Prog. Part. Nucl. Phys. {\bf 67} (2012) 1019.



\bibitem{LCQCDSR}
  P. Colangelo, F. De Fazio, G. Nardulli, N. Di Bartolomeo and R. Gatto, Phys. Rev. {\bf D52} (1995) 6422;
 T. M. Aliev, N. K. Pak and M. Savci, Phys. Lett. {\bf B390} (1997) 335;
 P. Colangelo and F. De Fazio, Eur. Phys. J. {\bf C4} (1998) 503;
 Y. B. Dai and S. L. Zhu, Phys. Rev. {\bf D58} (1998) 074009;
S. L. Zhu and Y. B. Dai, Phys. Rev. {\bf D58} (1998) 094033;
A. Khodjamirian, R. Ruckl, S. Weinzierl and O. I. Yakovlev, Phys. Lett. {\bf B457} (1999) 245;
 Z. H. Li, T. Huang, J. Z. Sun and Z. H. Dai, Phys. Rev. {\bf D65} (2002) 076005;
 H. c. Kim and S. H. Lee, Eur. Phys. J. {\bf C22} (2002) 707;
 D. Becirevic, J. Charles, A. LeYaouanc, L. Oliver, O. Pene and J. C. Raynal, JHEP {\bf 0301} (2003) 009;
Z. G. Wang and S. L. Wan, Phys. Rev. {\bf D73} (2006) 094020;
 Z. G. Wang and S. L. Wan, Phys. Rev. {\bf D74} (2006) 014017;
 Z. G. Wang, Eur. Phys. J. {\bf C52} (2007) 553;
Z. G. Wang, Nucl. Phys. {\bf A796} (2007) 61;
 Z. G. Wang, J. Phys. {\bf G34} (2007) 753;
Z. G. Wang, Phys. Rev. {\bf D77} (2008) 054024;
Z. G. Wang and Z. B. Wang, Chin. Phys. Lett. {\bf 25} (2008) 444;
Z. H. Li, W. Liu and H. Y. Liu, Phys. Lett. {\bf B659} (2008) 598.


\bibitem{SVZ79}  M. A. Shifman, A. I. Vainshtein and V. I. Zakharov, Nucl. Phys. {\bf B147} (1979) 385, 448.

\bibitem{Reinders85} L. J. Reinders, H. Rubinstein and S. Yazaki, Phys. Rept. {\bf 127} (1985) 1.

\bibitem{QCDSR-review}  P. Colangelo and A. Khodjamirian, arXiv:hep-ph/0010175.

\bibitem{Wang-BC} Z. G. Wang,  Eur. Phys. J. {\bf A49} (2013) 131.

\bibitem{Wang2013} Z. G. Wang,   Eur. Phys. J. {\bf C73} (2013) 2559.

\bibitem{Wang1209} Z. G. Wang,  Commun. Theor. Phys. {\bf 61} (2014) 81.

\bibitem{Gongshi-Ioffe} B .L. Ioffe and A. V. Smilga, Nucl. Phys. {\bf B216} (1983) 373;
 D. S. Du, J. W. Li and  M. Z. Yang, Eur. Phys. J. {\bf C37} (2004) 173.

\bibitem{Kiselev2000} V. V. Kiselev, A. K. Likhoded and A. I. Onishchenko, Nucl. Phys. {\bf B569} (2000) 473.

\bibitem{Coulomb-BC} V. V. Kiselev, Int. J. Mod. Phys. {\bf A11} (1996) 3689;
V. V. Kiselev, A. E. Kovalsky and A. K. Likhoded, Nucl. Phys. {\bf B585} (2000) 353.

\bibitem{Kiselev04}   V. V. Kiselev, Central Eur. J. Phys. {\bf 2} (2004) 523.

\bibitem{PDG} J. Beringer et al, Phys. Rev. {\bf D86} (2012) 010001.

\bibitem{Penin2004} A. A. Penin, A. Pineda, V. A. Smirnov and M. Steinhauser, Phys. Lett. {\bf B593} (2004) 124.


\bibitem{gg-conden} S. Narison, Phys. Lett. {\bf B693} (2010) 559; S. Narison, Phys. Lett. {\bf B706} (2012) 412;
S. Narison, Phys. Lett. {\bf B707} (2012) 259.


\end{thebibliography}
\end{document}